\documentclass[iop,apj,numberedappendix,appendixfloats,twocolappendix]{emulateapj}

\shorttitle{\sc Differential Kinematics}
\shortauthors{\sc Mathes et~al.}

\usepackage{natbib}
\usepackage{iondefs}
\usepackage{apjfonts}

\begin{document}

\title{Halo Mass Dependence of {\HI} and {\OVI} Absorption: Evidence for Differential Kinematics}

\author{
Nigel L. Mathes\altaffilmark{1},
Christopher W. Churchill\altaffilmark{1},
Glenn G. Kacprzak\altaffilmark{2,3},
Nikole M. Nielsen\altaffilmark{1}, \\
Sebastian Trujillo-Gomez\altaffilmark{1},
Jane Charlton\altaffilmark{4},
and
Sowgat Muzahid\altaffilmark{4}
}

\altaffiltext{1}{New Mexico State University, Las Cruces, NM 88003}
\altaffiltext{2}{Swinburne University of Technology, Victoria 3122, Australia}
\altaffiltext{3}{Australian Research Council Super Science Fellow}
\altaffiltext{4}{The Pennsylvania State University, University Park, PA 16802}

\begin{abstract}

We studied a sample of 14 galaxies ($0.1< z <0.7$) using {\it
  HST}/WFPC2 imaging and high-resolution {\it HST}/COS or {\it
  HST}/STIS quasar spectroscopy of {\Lya}, {\Lyb}, and {\OVIdblt}
absorption. The galaxies, having $10.8\leq \log(M_{\rm\,h}/M_{\odot})
\leq 12.2$, lie within $D=300$~kpc of quasar sightlines, probing out
to $D/R_{\rm vir}=3$.  When the full range of $M_{\rm\,h}$ and
$D/R_{\rm vir}$ of the sample are examined, $\sim\!40$\% of the {\HI}
absorbing clouds can be inferred to be escaping their host halo.  The
fraction of bound clouds decreases as $D/R_{\rm vir}$ increases such
that the escaping fraction is $\sim\!15$\% for $D/R_{\rm vir} < 1$,
$\sim\!45$\% for $1\leq D/R_{\rm vir} <2$, and $\sim\!90$\% for $2\leq
D/R_{\rm vir} < 3$.  Adopting the median mass $\log
M_{\rm\,h}/M_{\odot} = 11.5$ to divide the sample into ``higher'' and
``lower'' mass galaxies, we find mass dependency for the hot CGM
kinematics.  To our survey limits, {\OVI} absorption is found in only
$\sim\!40$\% of the {\HI} clouds in and around lower mass halos as
compared to $\sim\!85$\% around higher mass halos. For $D/R_{\rm vir}
<1$, lower mass halos have an escape fraction of $\sim\!65$\%, whereas
higher mass halos have an escape fraction of $\sim\!5$\%.  For $1\leq
D/R_{\rm vir} <2$, the escape fractions are $\sim\!55$\% and
$\sim\!35$\% for lower mass and higher mass halos, respectively.  For
$2\leq D/R_{\rm vir} <3$, the escape fraction for lower mass halos is
$\sim\!90$\%.  We show that it is highly likely that the absorbing
clouds reside within $4R_{\rm vir}$ of their host galaxies and that
the kinematics are dominated by outflows.  Our finding of
``differential kinematics'' is consistent with the scenario of
``differential wind recycling'' proposed by Oppenheimer {\etal} We
discuss the implications for galaxy evolution, the stellar to halo
mass function, and the mass metallicity relationship of galaxies.

\end{abstract}


%============== INTRODUCTION =============================

\section{Introduction}
\label{sec:intro}

Characterizing the baryonic gas processes within and surrounding
galaxies is central to understanding their formation and evolution.
Quantifying the spatial extent, kinematics, and, in particular, the
recycling and/or escape fraction of circumgalactic gas are of primary
importance in that they place direct observational constraints on
simulations of galaxies and provide insights into the workings of
galaxy evolution.

High-resolution spectroscopy of quasars, which provide a background
luminous source, and high-resolution imaging of the foreground
galaxies provides the data necessary for examining the kinematics of
galactic gas and its geometric distribution with respect to the galaxy
projected orientation.

In general, the gas structures in and around galaxies can be divided
into three broad categories: the interstellar medium (ISM), the
circumgalactic medium (CGM), and the intergalactic medium (IGM). The
CGM, being the gas reservoir that interfaces with the star-forming
ISM, outflowing stellar-driven winds, and the accreting IGM, may
contain up to 50\% of the baryonic mass bound to galaxies
\citep{Tumlinson2011} and account for up to 50\% of the baryons
unaccounted for in galaxy dark matter halos \citep{Werk2014}.  As
such, the CGM may play the most critical role in governing the
properties of galaxies \citep[e.g.,][]{Oppenheimer2010, MAGIICAT3},
including regulatory physics leading to the observed stellar mass
function \citep[e.g.,][]{Behroozi2013} and the stellar mass-ISM
metallicity relationship \citep[e.g.,][]{Tremonti2004}.

The physical extent of the CGM and the transition zone between the
CGM and IGM are currently open questions. Studies by
\citet{Steidel2010}, \citet{Prochaska2011}, and \citet{Rudie2012}
indicate a transition from the CGM to the IGM at $\log N({\HI}) \simeq
14$ and a projected distance of $\sim\! 300$~kpc from galaxies at
$z\sim 2.5$.  At this redshift, $\log N({\HI}) \simeq 14$ corresponds
to an overdensity of $\log \rho_{\hbox{\tiny
    H}}/\bar{\rho}_{\hbox{\tiny H}} \simeq 0.5$, whereas at $z\simeq
0$, this overdensity would suggest a CGM/IGM transition at $\log
N({\HI}) \simeq 13$ \citep[see][]{Dave1999}.  Indeed,
\citet{Ford2013mass} show that, in the over-dense regions hosting
galaxies, the extent of the {\HI} at fixed column density increases
with virial mass, suggesting that a single fiducial physical size for
the CGM may not apply across the entire mass range of galaxies; it may
be more appropriate to scale CGM properties relative to the virial
radius \citep[e.g.,][]{CWC2013Masses,MAGIICAT3}

%Mass 

Cosmological hydrodynamic simulations indicate that the virial mass
may dictate the temperature history, density, and mode of IGM
accretion \citep{Birnboim2003, Keres2005, Dekel2006, Keres2009,
  vandeVoort2011, Fumagalli2011, vandeVoort2012}.  The recycling
timescale of wind material back into the ISM may depend upon galaxy
virial mass according to what \citet{Oppenheimer2010} call
``differential wind recycling''.  Their simulations suggest that the
recycling time of wind material through the CGM could be shorter for
higher mass halos, and possibly longer than the Hubble time for the
lowest mass halos, and that this behavior may be key for understanding
the galaxy stellar mass function in the low mass range.  As deduced
from the simulations, differential wind recycling is primarily due to
greater hydrodynamic (not gravitational) deceleration of wind material
in higher mass halos due to their being embedded in denser gas
environments, leading to diminished recycling times \citep[also
  see][]{Oppenheimer2008}.

%D/Rvir 

As probed by {\MgII} absorption \citep[see][and references
  therein]{MAGIICAT2, MAGIICAT1}, the observed projected absorption
profile, covering fraction, and physical extent of the cool gas
component of the CGM combine to suggest that the cool/warm CGM
exhibits a self-similar radial behavior with virial mass
\citep{CWC2013Masses, MAGIICAT3}.  {\MgII} absorption properties
behave self-similarly with $D/R_{\rm vir}$, the projected distance of
the absorption from the galaxy relative to the virial radius.
\citet{Stocke2013} also report behavior that can be interpreted as
self-similarity in that, once impact parameter is scaled by $R_{\rm
  vir}$, the cool/warm CGM gas properties show little-to-no variation
as a function of projected distance, gas kinematics, and galaxy
luminosity \citep[however, see][who report weak anti-correlations in
  the cloud hydrogen number densities and ionization
  parameters]{Werk2014}.  If virial mass influences the global physics
of the CGM, then the CGM/IGM boundary and ISM/stellar formation
physics of galaxies may fundamentally be related to the dark matter
overdensity profile of halos within the virial radius.

%Orientation 

Simulations of starbursts predict that outflowing gas will
preferentially escape along the galaxy minor axis
\citep[e.g.,][]{Strickland2004}.  Infalling gas is predicted to
preferentially accrete in the galactic plane and kinematically trace
galaxy rotation \citep[e.g.,][]{Stewart2011}, consistent with the
observations of \citet{Steidel2002} and \citet{Kacprzak2010}.
Consistent with these predictions, {\MgII} absorption is most commonly
found along the projected minor and major axes of galaxies
\citep{Bouche2012, Kacprzak2012-PA}. \citet{Bordoloi2011} report that,
on average, larger {\MgII} equivalent widths are found along the
projected minor axis as compared to the projected major axis. The
distribution of {\HI}, traced using {\Lya} absorption, appears to be
more uniform with respect to the galaxy projected axis
\citep[e.g.,][]{Stocke2013}.

%Kinematics

The kinematics of CGM/IGM absorption with respect to host galaxy
escape velocity places direct observational constraints on the influence
of hydrodynamic and/or gravitational deceleration of CGM gas and
provides insights into the plausibility of a mass dependent wind
recycling scenario. For $D<150$~kpc (corresponding to $D/R_{\rm vir} <
1$), \citet{Tumlinson2011,Tumlinson2013} find that, in moderate to
high mass halos with $\log (M_{\rm\,h}/M_{\odot}) > 11.3$, CGM gas is
predominantly bound. \citet{Stocke2013} find that the majority of the
CGM within the projected virial radius also appears bound, but outside
of the projected virial radius, velocities can exceed the
escape velocity.

% Hot phase OVI 

Simulations, such as those by \citet{Keres2005} and by
\citet{vandeVoort2012}, predict that more massive halos have higher
hot gas mass fractions.  The metal-enriched hot phase of the CGM,
which can be traced by {\OVI} absorption, may be a reservoir of a
significant baryonic mass \citep{Tumlinson2011}.  The hot phase may
serve as a ``coronal'' hydrostatic region surrounding galaxies that
strongly governs the formation and destruction of the cool/warm CGM
``clouds'' \citep{Mo1996, Maller2004, Dekel2006}.  On the other hand,
gas traced by {\OVI} absorption may arise in multi-phase gas
structures \citep{Prochaska2004, Cooksey2008}.

Although {\OVI} absorbing gas has been extensively studied in the
Galactic, extragalactic, and IGM environments \citep{Savage2002,
  Savage2003, Richter2004, Sembach2004, Lehner2006, Danforth2008,
  Tripp2008, Thom2011, Tumlinson2011, Muzahid2014}, the potential
important role of CGM gas beckons further exploration of {\OVI}
absorption around galaxies in order to address basic questions such
as: what is the physical extent of {\OVI} absorbing gas around
galaxies, and where is the CGM/IGM transition region of the hot phase?
Can a CGM/IGM transition region, or boundary, be observationally
discerned?  Is the transition region halo mass dependent?  Are there
trends in the hot phase with respect to a galaxy's virial radius?
Does the hot phase have a preferred geometrical distribution around
galaxies, similar or dissimilar to what is seen for the cold/warm
phase?  Is there kinematic evidence for the differential wind
recycling scenario?

%Paper Outline 

In order to address these questions, we examine the kinematics and
spatial distribution of {\HI} and {\OVI} column densities in the
CGM/IGM and their absorption kinematics for a small sample of
galaxies.  This paper is structured as follows: In
Section~\ref{sec:sample} we discuss the sample selection, the data,
and data analysis. In Section~\ref{sec:CGM}, we examine the spatial
extent and geometry of the {\HI} and {\OVI} absorbing gas.  In
Section~\ref{sec:kin}, we compare the {\HI} and {\OVI} kinematics and
examine the spatial and virial mass dependence of the CGM with respect
to halo escape velocity. In Section~\ref{sec:discussion}, we discuss
our findings and in Section~\ref{sec:conclusion} we summarize our
results and discussion.  Throughout, we adopt a $\Lambda$CDM
cosmological model with $h=0.7$, where
$h=H_0/100~\hbox{{\kms}~Mpc$^{-1}$}$, with $\Omega_m=0.3$ and
$\Omega_{\Lambda}=0.7$.

%============== SAMPLE DESCRIPTION, DATA, ANALYSIS =======================

\section{Sample Selection, Data, and Analysis}
\label{sec:sample}

\subsection{Sample Selection}

We have assembled a sample of 14 galaxies in the fields of UV bright
quasars with high resolution {\it HST\/} imaging and ultraviolet
spectra.  We impose four primary criteria for the galaxy sample: (1)
each galaxy must be intervening to a background quasar within a
projected distance of 300 kpc from the line of sight and must have a
spectroscopic redshift measurement, (2) each galaxy must be imaged
with the {\it Hubble Space Telescope} ({\it HST\/}), (3) a {\it
  HST\/}/COS and/or STIS spectrum of the background quasar is
available that, at a minimum, covers the redshifted {\Lya}, {\Lyb}, and
{\OVIdblt} transitions within $\pm 1000$ {\kms} of the associated
foreground galaxy redshift, and (4) the foreground galaxy must not
reside in a group or cluster environment to the extent that the data
provides such information.

The 300 kpc projected distance allows us to study the CGM out to the
extent probed by \citet{Steidel2010}, \citet{Prochaska2011}, and
\citet{Rudie2012} and to extend beyond the 150 kpc range probed by
\citet{Tumlinson2011, Tumlinson2013}.  The {\it HST\/}/WFPC2 images
provide the spatial information required to measure galaxy
morphological parameters and determine galaxy orientations relative to
the quasar line of sight.  The {\it HST\/}/COS and/or STIS
spectroscopy allows Voigt profile decomposition of the {\Lya}, {\Lyb},
and {\OVIdblt} absorption profiles, providing individual ``cloud''
column densities and kinematics.

By selecting isolated field galaxies, we aim to study the CGM
independent of galaxy environment.  In a given quasar field, we first
require that no other galaxy is identified within $\pm 1000~{\kms}$
(based upon redshift). If other galaxies exist in the field within
this velocity window, we then require the galaxies lie farther than a
projected distance of $600~\mathrm{kpc}$ from the quasar line of
sight.

The quasar fields from which the galaxy sample is drawn were surveyed
by \citet{Ellingson1994}, \citet{Lanzetta1995}, \citet{LeBrun1996},
\citet{Chen2001}, and \citet{Johnson2013}.  We note that the fields
have been studied for different science goals employing different
facilities to varying degrees of completeness.  A detailed discussion
of the application of the galaxy selection criteria for each field is
presented in Appendix~\ref{sec:individualfields}.  Here, we briefly
summarize the surveys.  The galaxies observed by \citet{Ellingson1994}
have redshifts $0.3 < z < 0.6$ and are intervening to radio-selected
quasars.  The galaxies surveyed by \citet{Lanzetta1995},
\citet{LeBrun1996}, and \citet{Chen2001} have redshifts $0.05 < z <
0.8$ are selected on the basis that {\it HST\/}/FOS spectra of the
quasar had been obtained for the {\it HST\/} Key Project
\citep[cf.,][]{Bahcall1993}.

The resulting redshift range of our galaxy sample is determined
exclusively by the UV spectral coverage of the {\it HST\/}/COS and
STIS observations and not by any {\it a priori\/} redshift cuts.
Using the above four selection criteria, we compiled a sample of 14
galaxies spanning the redshift range of $0.12 \le z \le 0.67$ with
impact parameters from $60 \le D \le 290$ kpc.

% Observations Table =======================================

\begin{deluxetable}{lccrr}
\tablecolumns{5}
\tablewidth{0pt}
\setlength{\tabcolsep}{0.06in}
\tablecaption{Journal of Observations \label{tab:obs}}
\tablehead{
  \colhead{(1)} &
  \colhead{(2)} &
  \colhead{(3)} &
  \colhead{(4)} &
  \colhead{(5)} \\[2pt]
  \colhead{Quasar} &
  \colhead{Instrument} &
  \colhead{Filter/Grating} &
  \colhead{Exp. Time} &
  \colhead{PID} \\[1pt]
  \colhead{} &
  \colhead{} &
  \colhead{} &
  \colhead{[s]} & 
  \colhead{} }
\startdata
Q0405$-$123 & $HST$/WFPC2 & F702W & 2400 & 5949 \\[3pt]
            & $HST$/COS & G130M+G160M  & 20,749 & 11541 \\[3pt]
Q0454$-$2203 & $HST$/WFPC2 & F702W & 1200 & 5098 \\[3pt]
             & $HST$/COS & G160M & 2778 & 12252 \\[3pt]
             & $HST$/COS & G160M & 1849 & 12466 \\[3pt]
             & $HST$/COS & G185M & 74,410 & 12536 \\[3pt]
Q1001$+$2910 & $HST$/WFPC2 & F702W & 2400 &  5949 \\[3pt]
             & $HST$/COS & G130M+G160M & 12,988 & 12,038 \\[3pt]
Q1136$-$1334 & $HST$/WFPC2 & F702W & 2100 & 6919 \\[3pt]
             & $HST$/COS & G130M & 7751 & 12275 \\[3pt]
Q1216$+$0655 & $HST$/WFPC2 & F702W & 2100 & 6619 \\[3pt]
             & $HST$/COS & G130M+G160M & 10,702 & 12025 \\[3pt]
Q1259$+$5920 & $HST$/WFPC2 & F702W & 2100 & 6919 \\[3pt]
             & $HST$/WFPC2 & F702W & 2100 & 6919 \\[3pt]
             & $HST$/COS & G130M+G160M & 20,383 & 11541 \\[3pt]
Q1317$+$2743 & $HST$/WFPC2 & F702W & 4700 & 5984 \\[3pt]
             & $HST$/COS & G160M+G185M & 22,971 & 11667 \\[3pt]
Q1704$+$6048 & $HST$/WFPC2 & F702W & 2400 & 5949 \\[3pt]
             & $HST$/STIS & E140M & 22,155 & 8015 \\[3pt]
\\[-18pt]
\enddata
\end{deluxetable}

In Table~\ref{tab:obs}, we list the observational data employed for
the sample galaxies. Column (1) lists the quasar field [B1950
  designation]. Column (2) lists the imaging and spectroscopic
instruments. Column (3) lists the imaging filter and the COS or STIS
grating. Columns (4) and (5) list the exposure time and the program
ID, respectively.

% ============== GALAXY IMAGE DATA ======================================

\subsection{Galaxy Imaging and Photometric Properties}
\label{sec:imaging}

All {\it HST\/}/WFPC2 images were obtained using the F702W band. We
adopted the reduced and calibrated images from the WFPC-2
Associations Science Products Pipeline
(WASPP\footnote{http://archive.stsci.edu/hst/wfpc2/pipeline.html}).

The galaxy apparent Vega magnitudes, $m_{\hbox{\tiny F702W}}$, were
determined using $1.5~\sigma$ isophotes from Source Extractor
\citep{bertin1996}.  From the galaxy centroids, we compute the galaxy
offset from the quasar (arcsec) and the galaxy-quasar sightline impact
parameter (kpc).  We computed AB $r$-band absolute magnitudes, $M_r$,
by $k$-correcting the observed F702W magnitudes following the method
of \citet{MAGIICAT1}.  To determine galaxy virial masses,
$M_{\rm\,h}$, we performed halo abundance matching
\citep[e.g.,][]{Behroozi2010, Trujillo2011} following the method of
\citet{MAGIICAT3}, in which we match the distribution of the maximum
circular velocity of halos in the Bolshoi $N$-body cosmological
simulation of \citet{Klypin2011} to the COMBO-17 $r$-band luminosity
function of \citet{Wolf2003}.  Galaxy virial radii are then computed
from $M_{\rm\,h}$ using the relation of \citet{BryanNorman1998}.
Uncertainties in the viral masses and virial radii, which are on the
order 10\%, originate from the scatter in the virial mass circular
velocity distribution function \citep[see][for details]{MAGIICAT3}.

Quantified galaxy morphological parameters were measured using GIM2D
\citep{Simard2002} following the methods of \citet{Kacprzak2011}.
GIM2D models the two-dimensional brightness profiles of the galaxies
and computes the inclination, $i$, and the position angle, $\Phi$, on
the sky.  We adopt the formalism that $i=0^{\circ}$ is face-on and
$i=90^{\circ}$ is edge-on.  We translate the position angle to an
``azimuthal angle'' defined such that for $\Phi = 0^{\circ}$ the
quasar sightline lies along the projected major axis, and for $\Phi =
90^{\circ}$ it lies along the galaxy projected minor axis.

In Table~\ref{tab:galdata}, columns (1) and (2) list the quasar field
and the galaxy spectroscopic redshift. Columns (3)--(5) list
the galaxy offsets relative to the quasar and the galaxy impact
parameter, respectively. Columns (6) and (7) list the galaxy {\it
  HST\/}/WFPC2 F702W apparent magnitude (Vega) and the $r$-band
absolute magnitude (AB). Columns (8) and (9) list the virial mass and
virial radius of the galaxy. Columns (10) and (11) list the galaxy
azimuthal angle and inclination.

The galaxy WFPC/F702W apparent magnitudes range from $22.6 \ge
m_{\hbox{\tiny F702W}} \ge 17.8$. Absolute $r$-band magnitudes range
from $-15.5 \ge M_{r} \ge -20.2$.  The range of galaxy inclinations
and azimuthal angles are $18^{\circ} \le i \le 85^{\circ}$ and
$6^{\circ} \le \Phi \le 87^{\circ}$, respectively.  The virial masses
range from $10.8 \le \log(M_{\rm\,h}/M_{\odot}) \le 12.2$, with virial
radii between $70 \le R_{\rm vir} \le 225~\mathrm{kpc}$.  The median
virial mass is $\log(M_{\rm\,h}/M_{\odot}) = 11.5$.

% Galaxy Table =================================================

\begin{deluxetable*}{lcrrrcccccc}
\tablecolumns{11}
\tablewidth{0pt}
\setlength{\tabcolsep}{0.06in}
\tablecaption{Galaxy Properties \label{tab:galdata}}
\tablehead{
  \colhead{(1)} &
  \colhead{(2)} &  
  \colhead{(3)} &  
  \colhead{(4)} &  
  \colhead{(5)} &  
  \colhead{(6)} &  
  \colhead{(7)} &  
  \colhead{(8)} &  
  \colhead{(9)} &  
  \colhead{(10)} &
  \colhead{(11)} \\[2pt]
  \colhead{Quasar}  &
  \colhead{$z_{\rm gal}$} & 
  \colhead{$\Delta\alpha$} & 
  \colhead{$\Delta\delta$} & 
  \colhead{$D$} &
  \colhead{$m_{\rm\,F702W}$} & 
  \colhead{$M_{r}$} & 
  \colhead{$\log(M_{\rm\,h}$)} & 
  \colhead{$R_{\rm vir}$} & 
  \colhead{$\Phi$} & 
  \colhead{$i$} \\[1pt]
  \colhead{} &
  \colhead{} &
  \colhead{[arcsec]} & 
  \colhead{[arcsec]} & 
  \colhead{[kpc]} &
  \colhead{[Vega]} & 
  \colhead{[AB]} &
  \colhead{[M$_\odot$]} & 
  \colhead{[kpc]} & 
  \colhead{[deg]} & 
  \colhead{[deg]} }
\startdata
Q0405$-$123 & 0.1534 & $-66.2$ & $-30.9$ & 196 & 18.19 & $-19.21$ & 11.8$^{+0.4}_{-0.2}$ & 152$^{+56}_{-23}$ & 26.3$^{+0.9}_{-1.0}$ & 49.5$^{+0.5}_{-0.7}$ \\[3pt]
Q0405$-$123 & 0.2978 & $31.9$ & $-47.9$ & 258 & 19.23 & $-19.86$ & 12.2$^{+0.2}_{-0.2}$ & 224$^{+41}_{-28}$ & 22.4$^{+1.1}_{-1.3}$ & 62.1$^{+1.9}_{-2.8}$ \\[3pt]
Q0405$-$123 & 0.4100 & $2.6$ & $-23.2$ & 292 & 22.58 & $-17.37$ & 11.2$^{+0.5}_{-0.2}$ & 106$^{+52}_{-15}$ & 44.4$^{+28.0}_{-46.0}$ & 60.7$^{+24.3}_{-37.6}$ \\[3pt]
Q0454$-$2203 & 0.3818 & $0.3$ & $-19.7$ & 103 & 20.41 & $-19.35$ & 12.0$^{+0.3}_{-0.2}$ & 194$^{+46}_{-26}$ & 63.8$^{+4.3}_{-2.7}$ & 57.1$^{+19.9}_{-2.4}$ \\[3pt]
Q1001$+$2910 & 0.1380 & $-3.4$ & $-23.1$ &  57 & 21.65 & $-15.49$ & 10.8$^{+0.7}_{-0.2}$ & 73$^{+48}_{-11}$ & 12.4$^{+2.4}_{-2.9}$ & 79.1$^{+2.2}_{-2.1}$ \\[3pt]
Q1001$+$2910 & 0.2143 & $13.4$ & $-61.8$ & 222 & 21.64 & $-16.59$ & 11.2$^{+0.6}_{-0.2}$ & 98$^{+55}_{-15}$ & 14.2$^{+44.2}_{-42.7}$ & 18.1$^{+20.5}_{-18.1}$ \\[3pt]
Q1136$-$1334 & 0.1755 & $-3.2$ & $55.7$ & 166 & 21.30 & $-16.43$ & 11.0$^{+0.7}_{-0.2}$ & 86$^{+56}_{-13}$ & 44.3$^{+4.5}_{-4.9}$ & 84.8$^{+0.2}_{-3.4}$ \\[3pt]
Q1136$-$1334 & 0.2044 & $10.8$ & $-25.5$ &  93 & 19.69 & $-18.42$ & 11.7$^{+0.4}_{-0.2}$ & 146$^{+53}_{-21}$ & 5.8$^{+0.4}_{-0.5}$ & 83.4$^{+0.4}_{-0.5}$ \\[3pt]
Q1216$+$0655 & 0.1242 & $37.2$ & $-18.8$ &  95 & 17.78 & $-19.11$ & 11.7$^{+0.4}_{-0.2}$ & 146$^{+56}_{-22}$ & 68.0$^{+0.4}_{-0.5}$ & 85.0$^{+0.0}_{-0.0}$ \\[3pt]
Q1259$+$5920 & 0.1967 & $27.0$ & $-31.3$ & 135 & 20.55 & $-17.46$ & 11.2$^{+0.6}_{-0.2}$ & 103$^{+62}_{-16}$ & 39.7$^{+2.8}_{-2.2}$ & 80.7$^{+4.3}_{-3.2}$ \\[3pt]
Q1259$+$5920 & 0.2412 & $-23.4$ & $68.5$ & 280 & 19.58 & $-18.96$ & 11.9$^{+0.3}_{-0.2}$ & 169$^{+50}_{-24}$ & 42.5$^{+4.0}_{-3.7}$ & 71.9$^{+1.5}_{-2.9}$ \\[3pt]
Q1317$+$2743 & 0.6610 & $10.2$ & $-7.4$ & 103 & 21.34 & $-20.15$ & 12.1$^{+0.2}_{-0.2}$ & 224$^{+35}_{-25}$ & 87.0$^{+1.0}_{-1.0}$ & 65.8$^{+1.2}_{-1.2}$ \\[3pt]
Q1704$+$6048 & 0.1877 & $72.9$ & $5.9$ & 231 & 18.05 & $-19.85$ & 12.0$^{+0.3}_{-0.2}$ & 190$^{+50}_{-27}$ & 47.1$^{+0.7}_{-0.8}$ & 60.9$^{+0.6}_{-0.6}$ \\[3pt]
Q1704$+$6048 & 0.3380 & $-31.3$ & $-9.4$ & 159 & 21.26 & $-18.16$ & 11.6$^{+0.4}_{-0.2}$ & 140$^{+53}_{-21}$ & 53.8$^{+3.6}_{-2.9}$ & 53.1$^{+7.1}_{-15.3}$ \\[3pt]
\\[-18pt]
\enddata
\end{deluxetable*}

% Absorption Table =========================================

\begin{deluxetable*}{lcccccccrr}
\tablecolumns{10}
\tablewidth{0pt}
\setlength{\tabcolsep}{0.06in}
\tablecaption{Absorption Properties \label{tab:absdata}}
\tablehead{
  \colhead{(1)} &
  \colhead{(2)} &  
  \colhead{(3)} &  
  \colhead{(4)} &  
  \colhead{(5)} &  
  \colhead{(6)} &  
  \colhead{(7)} &  
  \colhead{(8)} &
  \colhead{(9)\tablenotemark{a}} &
  \colhead{(10)\tablenotemark{a}} \\[2pt]
  \colhead{Quasar}  &
  \colhead{$z_{\rm gal}$} &
  \colhead{$z_{\rm abs}$} & 
  \colhead{$W_r({\Lya})$} & 
  \colhead{$W_r({\Lyb})$} & 
  \colhead{$W_r({\OVI})$} & 
  \colhead{$\log N(\HI)$} & 
  \colhead{$\log N(\OVI)$} & 
  \colhead{$v^{\,(-)}$} & 
  \colhead{$v^{\,(+)}$} \\[1pt]
  \colhead{} &
  \colhead{} &
  \colhead{} &
  \colhead{[\AA]} &
  \colhead{[\AA]} &
  \colhead{[\AA]} &
  \colhead{[\cmsq]} &
  \colhead{[\cmsq]} &
  \colhead{[\kms]} &
  \colhead{[\kms]} }
\startdata
Q0405$-$123 & 0.1534 & 0.1530 & 0.547 $\!\pm\!$ 0.014 & 0.112 $\!\pm\!$ 0.011 & 0.019 $\!\pm\!$ 0.004 & 14.13 $\!\pm\!$ 0.03 & 13.40 $\!\pm\!$ 0.07 & $-575$ & $212$ \\[3pt]
Q0405$-$123 & 0.2978 & 0.2977 & 0.343 $\!\pm\!$ 0.014 & 0.061 $\!\pm\!$ 0.007 & 0.036 $\!\pm\!$ 0.006 & 14.00 $\!\pm\!$ 0.06 & 13.50 $\!\pm\!$ 0.23 & $-194$ & $87$ \\[3pt]
Q0405$-$123 & 0.4100 & 0.4059 & 0.966 $\!\pm\!$ 0.021 & 0.437 $\!\pm\!$ 0.011 & 0.048 $\!\pm\!$ 0.007 & 14.98 $\!\pm\!$ 0.95 & 13.76 $\!\pm\!$ 0.11 & $-1024$ & $-48$ \\[3pt]
Q0454$-$2203 & 0.3818 & 0.3817 & 0.609 $\!\pm\!$ 0.131 & 0.368 $\!\pm\!$ 0.052 & 0.255 $\!\pm\!$ 0.059 & 14.63 $\!\pm\!$ 0.25 & 14.41 $\!\pm\!$ 0.65 & $-230$ & $248$ \\[3pt]
Q1001$+$2910 & 0.1380 & 0.1375 & 0.776 $\!\pm\!$ 0.018 & 0.322 $\!\pm\!$ 0.017 & 0.084 $\!\pm\!$ 0.010 & 14.91 $\!\pm\!$ 0.06 & 14.03 $\!\pm\!$ 0.03 & $-768$ & $12$ \\[3pt]
Q1001$+$2910 & 0.2143 & 0.2128 & 0.736 $\!\pm\!$ 0.055 & 0.285 $\!\pm\!$ 0.037 & 0.073 $\!\pm\!$ 0.014 & 14.33 $\!\pm\!$ 0.33 & 14.00 $\!\pm\!$ 0.13 & $-556$ & $50$ \\[3pt]
Q1136$-$1334 & 0.1755 & 0.1749 & 0.661 $\!\pm\!$ 0.050 & $<$ 0.227 & $<$ 0.356 & 14.35 $\!\pm\!$ 0.10 & $<$ 13.50 & $-347$ & $49$ \\[3pt]
Q1136$-$1334 & 0.2044 & 0.2044 & 1.383 $\!\pm\!$ 0.173 & 1.395 $\!\pm\!$ 0.033 & 0.147 $\!\pm\!$ 0.033 & 15.94 $\!\pm\!$ 0.85 & 14.29 $\!\pm\!$ 0.16 & $-456$ & $293$ \\[3pt]
Q1216$+$0655 & 0.1242 & 0.1242 & 1.376 $\!\pm\!$ 0.022 & 0.660 $\!\pm\!$ 0.063 & 0.447 $\!\pm\!$ 0.058 & 15.26 $\!\pm\!$ 0.38 & 14.72 $\!\pm\!$ 0.26 & $-254$ & $268$ \\[3pt]
Q1259$+$5920 & 0.1967 & 0.1963 & 0.442 $\!\pm\!$ 0.024 & $<$ 0.123 & 0.040 $\!\pm\!$ 0.004 & 13.99 $\!\pm\!$ 0.15 & 13.73 $\!\pm\!$ 0.24 & $-307$ & $441$ \\[3pt]
Q1259$+$5920 & 0.2412 & 0.2412 & 0.049 $\!\pm\!$ 0.015 & $<$ 0.042 & $<$ 0.088 & 13.06 $\!\pm\!$ 0.07 & $<$ 11.84 & $-100$ & $99$ \\[3pt]
Q1317$+$2743 & 0.6610 & 0.6605 & 1.542 $\!\pm\!$ 0.091 & 0.905 $\!\pm\!$ 0.030 & 0.258 $\!\pm\!$ 0.043 & 18.53 $\!\pm\!$ 0.87 & 14.51 $\!\pm\!$ 0.08 & $-514$ & $344$ \\[3pt]
Q1704$+$6048 & 0.1877 & 0.1875 & 0.387 $\!\pm\!$ 0.011 & 0.044 $\!\pm\!$ 0.007 & 0.053 $\!\pm\!$ 0.013 & 14.08 $\!\pm\!$ 0.01 & 13.72 $\!\pm\!$ 0.07 & $-133$ & $80$ \\[3pt]
Q1704$+$6048 & 0.3380 & 0.3380 & $<$ 0.086 & $<$ 0.049 & $<$ 0.050 & $<$ 12.00 & $<$ 12.00 & $-100$ & $99$ \\[3pt]
\\[-18pt]
\enddata
\tablenotetext{a}{The velocities correspond to the {\Lya} absorption.}
\end{deluxetable*}

% ============= QUASAR ABSORPTION SPECTRA =====================================

\subsection{Quasar Spectra and Absorption Properties}
\label{sec:spectra}

The {\it HST\/}/COS spectra were reduced and flux calibrated using the
CalCOS pipeline (V2.11). Vacuum and heliocentric corrections,
dispersion alignment, and co-addition of individual exposures were
performed using software developed by the COS
team\footnote{http://casa.colorado.edu/$\!\sim\!$danforth/science/cos/costools.html}
\citep[also see][]{Narayanan2011}.

Reduction and calibration of the E140M {\it HST\/}/STIS spectrum (for
Q1704$+$6048 using the $0.2^{\prime\prime} \times 0.2^{\prime\prime}$
slit) was performed using the standard STIS pipeline
\citep{Brown2002}.  Further details are discussed in
\citet{Narayanan2005}. Continuum fitting for both the {\it HST\/}/COS
and {\it HST\/}/STIS data sets was conducted using the interactive
SFIT task in IRAF\footnote{IRAF is distributed by the National Optical
t  Astronomy Observatory, which is operated by the Association of
  Universities for Research in Astronomy (AURA) under cooperative
  agreement with the National Science Foundation.} following the
methods described in \citet{Sembach1992}. We then refined higher order
continuum fits using our own code, {\sc Fitter} \citep{Churchill2000}.

We searched the quasar spectra for {\Lya}, {\Lyb}, and {\OVIdblt}
absorption within $\pm1000$ {\kms} of the identified galaxy redshift.
We adopt the objective detection methods of \cite{Schneider1993} using
the $5~\sigma$ uncertainty in the equivalent width spectrum.  Once an
absorption feature is identified, we use the methods of
\citet{Churchill1999} and \citet{Churchill2001} to measure the
velocity extremes of the absorption, $v^{\,(-)}$ and $v^{\,(+)}$, the
rest-frame equivalent widths, $W_r$, and the optical depth mean system
absorption redshifts, $z_{\rm abs}$, using the {\Lya} absorption feature.

\begin{figure*}[bth]
\epsscale{1.15}
\plotone{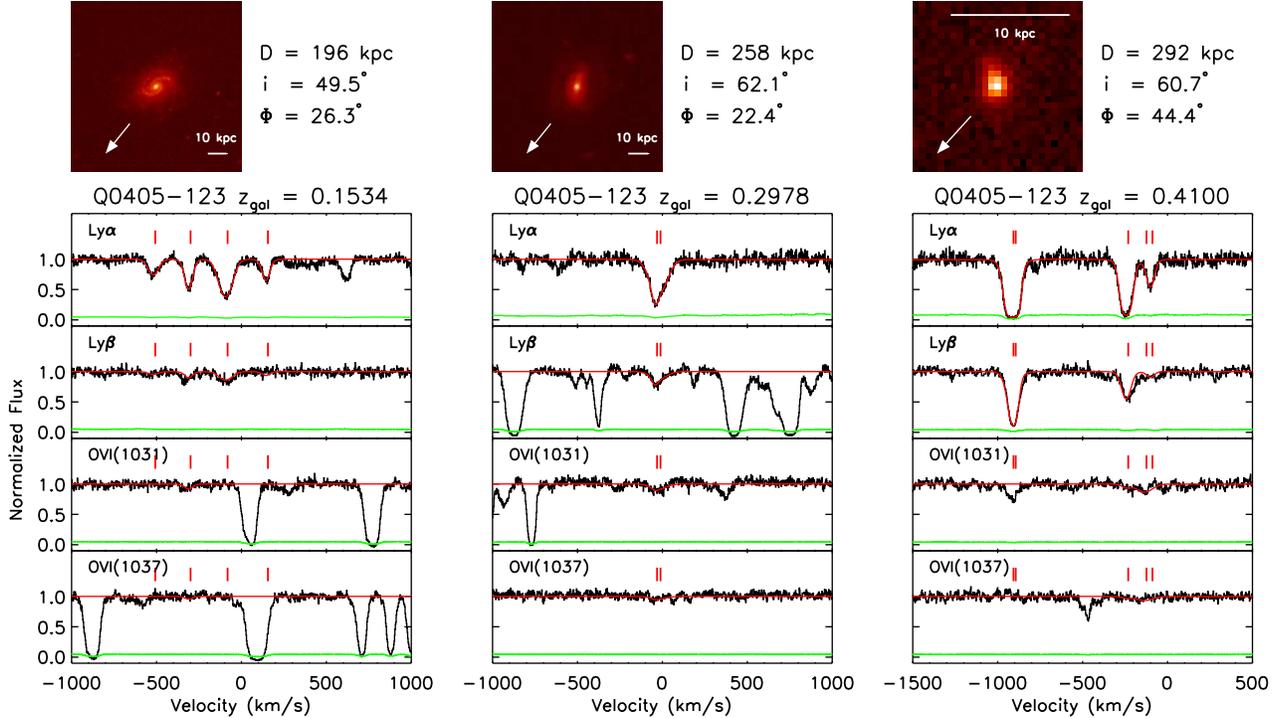}
\caption{Galaxy and absorption data for the galaxies at $z_{\rm gal} =
  0.1534$, $z_{\rm gal} = 0.2978$, and $z_{\rm gal} = 0.4100$ in the
  field toward Q0405$-$123.  We show the {\it HST\/}/WFPC2 image with
  an arrow pointing in the direction of the quasar sightline.  The
  galaxy impact parameter, inclination, and azimuthal angle are
  listed.  We also show the observed spectra and Voigt profile fits
  for the associated absorption. Plotted in black is the normalized,
  continuum-fitted UV spectrum and in green is the $1~\sigma$
  uncertainty spectrum. Overlaid in red is the VP model fit for each
  transition. Red ticks mark the centroid of each VP
  component. Velocities are rest-frame relative to the associated
  galaxy systemic velocity.}
\label{fig:profile}
\end{figure*}

For an absorption feature to be adopted as a detection, we apply the
criterion of a $3~\sigma$ equivalent width significance level, i.e.,
$W_r \geq 3\sigma_{W_r}$, otherwise we quote $3\sigma_{W_r}$ as the
upper limit on $W_r$.  The quoted uncertainties in the measured
equivalent widths account for both the pixel statistical uncertainty
and the systematic uncertainty due to the choice of continuum fit.
The latter assumes a mean continuum placement uncertainty of 30\% of
the mean pixel statistical uncertainty
\citep[see][]{Sembach1992}. Depending on the signal-to-noise ratio of
the spectral region, the continuum uncertainty yields a 5--20~m{\AA}
systematic uncertainty in the equivalent width.

To verify the identity of {\Lya} absorption, we examine whether the
associated {\Lyb} absorption is formally detected. Further
verification is obtained by detection of the {\OVI} doublet; however,
not all {\Lya} absorption has detected {\Lyb} and/or metal-line
absorption.  In the case of the {\OVI} doublet, we do not require that
both members of the doublet are formally detected in order to identify
either {\OVI}~$\lambda 1031$ or {\OVI}~$\lambda 1037$ absorption
(either due to line blending or the $\lambda 1037$ absorption being
below the required significance level).  The {\Lya} transition was
detected for all but one of the galaxy-absorber pairs.  For all but
four galaxy-absorber pairs, {\Lyb} was detected, but {\OVI} was
detected for three of these four pairs.  For only one pair, no {\Lya},
{\Lyb}, nor {\OVI} was detected.

To quantify the absorption column densities and kinematics, we fitted
the absorption line using Voigt profile (VP) decomposition.  Each VP
component is described by three physical parameters, the column
density, Doppler $b$ parameter, and the velocity center.  We employed
the code {\sc Minfit} \citep{Churchill2001}, adopting the philosophy
of enforcing the minimum number of statistically required VP
components to model an absorption system (i.e., simultaneously to all
transitions).  Details of the fitting procedure are described in
\cite{EvansThesis}.  We tie the velocity centers for each VP component
across all ions.  We also assume the line broadening is dominated by a
Gaussian turbulent component; therefore, each VP component at a given
velocity has the same Doppler $b$ parameter.  The latter assumption
is motivated by the simulations of \citet{Oppenheimer2009} in which
sub-resolution turbulence was required for modeling {\OVI} absorbing
gas in order to reproduce the observed column density and $b$
parameter distributions.  The one exception is $z_{\rm gal} = 0.1963$
absorption in the spectrum of Q1259$+$5920, where a satisfactory VP
model required thermal scaling of the Doppler $b$ parameters.

Though {\Lyg} is detected for several, but not all of the absorption
systems, we quote VP model results only from simultaneous fits to the
{\Lya} and {\Lyb} transitions (except in three systems in which only
{\Lya} was detected).  \citet{Muzahid2014} showed that {\Lya} and
{\Lyb} absorption primarily traces the high ionization phase where
{\OVI} arises, whereas the higher order Lyman series line profiles are
dominated by the lower ionization phase giving rise to {\CII},
{\SiII}, {\CIII}, and {\SiIII} absorption.  Thus, by omitting higher
order Lyman series lines we do not lose information on the {\OVI}
phase.  Most importantly, by omitting higher order {\HI} Lyman series
lines, even when they are detected, we present a uniform analysis of
the absorption systems.  In Appendix~\ref{sec:Lyg}, we compare the
derived {\HI} column densities, $N({\HI})$, obtained from
{\Lya}+{\Lyb} VP models and {\Lya}+{\Lyb}+{\Lyg} VP models for systems
for which {\Lyg} is also detected.  The exercise demonstrates that the
inclusion of {\Lyg} does not discernibly alter our derived $N({\HI})$
values.

In the case of line blending (overlapping absorption from transitions
associated with systems at other redshifts), when possible, we
carefully decompose the lines using the procedure illustrated in
Appendix~\ref{sec:deblending}.  The deblending technique is designed
to recover the shape of the profile for the target transition.

Results of the absorption line analysis are listed in Table
\ref{tab:absdata}. Column (1) gives the quasar field. Columns (2) and
(3) list the spectroscopic galaxy redshift and the absorption
redshift. Columns (4)--(6) list the measured rest-frame equivalent
widths of the {\Lya}, {\Lyb}, and {\OVIfirst} absorption
profiles. Columns (7) and (8) give the system total {\HI} and {\OVI}
column densities, which are the sums of the VP components in each
system. Columns (9) and (10) list the maximum blueward and redward
velocity limits of the {\Lya} absorption profiles.

\begin{figure*}[thb]
\plotone{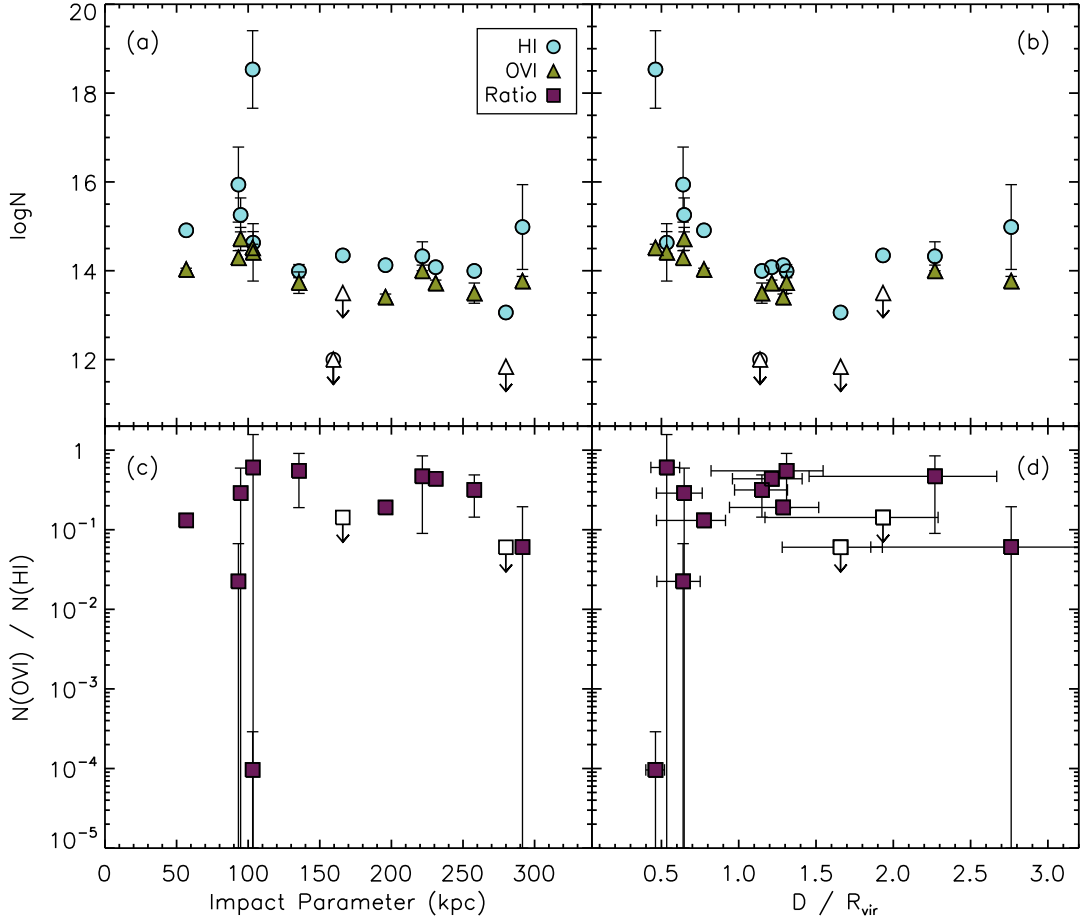}
\caption{The system total $N({\HI})$ [sky-blue points], $N({\OVI})$
  [green], and ratio $N({\OVI})/N({\HI})$ [magenta] plotted against
  $D$ [panels (a) and (c)] and against $D/R_{\rm vir}$ [panels (b) and
    (d)].  Open circles with downward arrows represent upper limits.
  The uncertainties in $D$ are less than 1 kpc.  The uncertainties in
  $D/R_{\rm vir}$ are shown in panel (f) only.}
\label{fig:DistancePanel}
\end{figure*}

The range of rest-frame equivalent widths is $0.05 \le W_r(\Lya) \le
1.54$ {\AA} and $0.02 \le W_r(1031) \le 0.45$ {\AA}, corresponding to
the system total column densities ranges $13.1 \le \log N({\HI}) \le
18.5$ and $13.4 \le \log N({\OVI}) \le 14.7$, respectively. The
maximum blueward and redward velocities of the {\Lya} absorption are
$v^{\,(-)} = -1020$~{\kms} and $v^{\,(+)} = 440$~{\kms}, respectively.

\subsection{Presentation of Galaxy-Absorber Pairs}

In Figure~\ref{fig:profile}, we show three of the 14 galaxy-absorber
pairs in our sample.  The remaining 11 galaxies and their associated
absorption are presented in Appendix~\ref{sec:individualfields}.

For each galaxy-absorber pair, we present a portion of the {\it
  HST\/}/WFPC2 image centered on the galaxy and the {\Lya}, {\Lyb},
and {\OVIdblt} absorption profiles.  In the galaxy image, the arrow
points in the direction of the quasar line of sight.  A bar provides
that scale of 10 kpc in the galaxy rest frame.  The legend provides
the galaxy impact parameter, $D$, inclination, $i$, and azimuthal
angle, $\Phi$.  The Voigt profile models of the absorption lines are
the red curves superimposed on the black data.  The velocity centroid
of each VP component is shown by the red ticks above the continuum
normalized spectra.  The velocity zero-point of each spectrum is taken
to be galaxy systemic velocity.

%============= RESULTS: CGM EXTENT AND GEOMETRY =========================

\section{CGM Extent and Geometry}
\label{sec:CGM}

% ================ DISTANCE RESULTS ================
\subsection{Spatial Behavior}
\label{sec:distance}

In Figure~\ref{fig:DistancePanel}, we present the spatial behavior of
the system total {\HI} column density, $N({\HI})$, and the system
total {\OVI} column density, $N({\OVI})$.  The system total column
densities are the sums of the individual Voigt profile component
column densities.  We color the $N({\HI})$ points sky-blue, the
$N({\OVI})$ points green, and the ratio $N({\OVI})/N({\HI})$ magenta.
Upper limits are shown as open circles with downward arrows.

In Figure~\ref{fig:DistancePanel}(a), we plot $N({\HI})$ and
$N({\OVI})$ versus impact parameter, $D$.  The total $N({\HI})$ is
typically $\log N({\HI}) = 14$ out to $D\sim300$~kpc (we note that one
system has a stringent upper limit of $\log N({\HI}) < 12$).  Though
higher $N({\HI})$ systems are found at $D<100$~kpc, there is no
statistical trend between $N({\HI})$ and $D$. A Kendall-$\tau$ rank
correlation test, which includes upper limits, yields a $1.8~\sigma$
consistency with the null hypothesis of no correlation.

We find $N({\OVI})$ out to $\sim\!290$~kpc to a limit of $\log
N({\OVI}) = 12.8$.  We note that the detection at $D=292$~kpc (in
Q0454$-$132 at $z=0.4100$) is very tentative (see
Figure~\ref{fig:profile}). If that detection were deemed an upper
limit, then our data show {\OVI} absorption out to $\sim\!260$~kpc.
Similar to the behavior with $N({\HI})$, we find no statistically
significant trend between $N({\OVI})$ and $D$ (only a $2.5~\sigma$
trend).  

\begin{figure}[th]
\epsscale{1.15}
\plotone{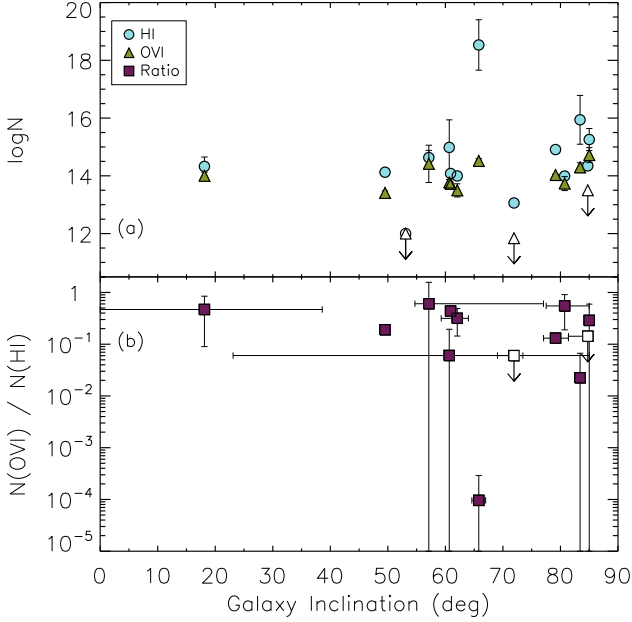}
\vglue 0.1in
\caption{(a) The system total $N({\HI})$ [sky-blue] and $N({\OVI})$
  [green] plotted against inclination $i$.  (b) The ratio
  $N({\OVI})/N({\HI})$ [magenta] plotted against inclination.  Open
  circles with downward arrows represent upper limits.  The
  uncertainties in the inclination are shown in panel (b) only.}
\label{fig:inclination}
\end{figure}

In Figure~\ref{fig:DistancePanel}(b), we plot $N({\HI})$ and
$N({\OVI})$ versus $D/R_{\rm vir}$.  There is a higher average value
and a broader spread in $N({\HI})$ for $D/R_{\rm vir} < 1$ as compared
to $D/R_{\rm vir} > 1$, with $ \log \langle N({\HI}) \rangle = 15.9
\pm 1.6$ inside the projected virial radius and $\log \langle N({\HI})
\rangle = 14.1 \pm 0.5$. outside the projected virial radius.  In We
find that the $N({\OVI})$ values inside and outside of the projected
virial radius lie within one standard deviation of the average $\log
\langle N({\OVI}) \rangle = 14.03 \pm 0.44$.  We do not see a larger
dispersion in $N({\OVI})$ at $D/R_{\rm vir} < 1$ as seen for {\HI}.

The ratio $N({\OVI})/N({\HI})$ is shown in
Figures~\ref{fig:DistancePanel}(c) and \ref{fig:DistancePanel}(d) as a
function of $D$ and $D/R_{\rm vir}$, respectively.  Due to the flat
spatial distribution of $N({\OVI})$ and the higher dispersion of
$N({\HI})$ at $D/R_{\rm vir} < 1$, the spread in $N({\OVI})/N({\HI})$
is a factor of $\simeq 8$ greater inside than outside the projected
virial radius, with $\sigma(D/R_{\rm vir}\!\leq\!1) = 1.5$ and
$\sigma(D/R_{\rm vir}\!>\!1) = 0.18$.

The different spreads in the $N({\OVI})/N({\HI})$ inside and outside
the projected virial radius are driven by the spatial behavior of
{\HI}.  Inside the projected virial radius, $N({\HI})$ has a larger
dispersion than outside this region, whereas $N({\OVI})$ has a small
dispersion both inside and outside the projected virial radius.  The
{\OVI} column densities inside the projected virial radius appear to
be similar to those outside this region, whereas the quantity of {\HI}
can be variable.

% ================ ORIENTATION RESULTS ================
\subsection{Galaxy Orientation}
\label{sec:PA}

To ensure we have a fair sample with which we can examine the
geometric distribution of {\HI} and {\OVI} absorption around the
galaxies, we performed Kolmogorov-Smirnov (KS) tests to determine
whether the observed distributions of galaxy azimuthal angles, $\Phi$,
and inclinations, $i$, are consistent with being drawn from the
distributions expected for an unbiased sample.  For $\Phi$, the KS
test probability is $P({\rm KS}) = 0.62$ and for $i$ the probability
is $P({\rm KS}) = 0.46$.  We conclude that both $\Phi$ and $i$ are
consistent with unbiased distributions.

\begin{figure}[tbh]
\epsscale{1.15} 
\plotone{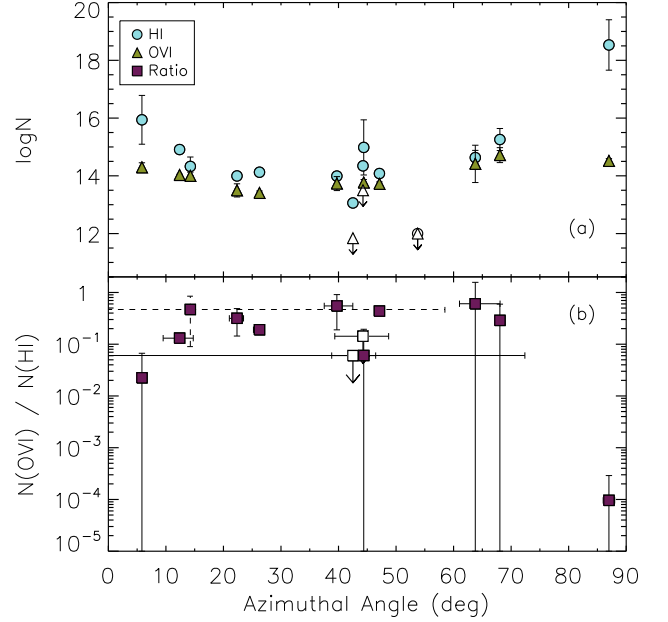}
\vglue 0.1in
\caption{(a) The system total $N({\HI})$ [sky-blue] and $N({\OVI})$
  [green] plotted against azimuthal angle, $\Phi$.  (b) The ratio
  $N({\OVI})/N({\HI})$ [magenta] plotted against azimuthal angle.
  Open circles with downward arrows represent upper limits.  The
  uncertainties in the azimuthal angle are shown in panel (b) only.}
\label{fig:PAPanel}
\end{figure}

In Figure~\ref{fig:inclination}(a) and \ref{fig:inclination}(b), we
show the system total $N({\HI})$, $N({\OVI})$, and
$N({\OVI})/N({\HI})$, respectively, versus inclination. The
uncertainties in the inclination measurements derived from GIM2D are
shown only in Figure~\ref{fig:inclination}(b).  A Kendall-$\tau$ test
(including limits) yields no trend in the $N({\HI})$ distribution
($1.1~\sigma$), the $N({\OVI})$ distribution ($0.8~\sigma$), nor in
the distribution of the ratio $N({\OVI}) / N({\HI})$ ($0.8~\sigma$)
with galaxy inclination. Whether a galaxy is observed with a face-on
or an edge-on orientation, there appears to be a relatively flat
distribution of {\HI} and {\OVI} column densities.

In Figure~\ref{fig:PAPanel}(a) and \ref{fig:PAPanel}(b), we present
the system total $N({\HI})$, $N({\OVI})$, and $N({\OVI})/N({\HI})$,
respectively, versus azimuthal angle, $\Phi$. The uncertainties in the
azimuthal angle measurements derived from GIM2D are shown only in
Figure~\ref{fig:PAPanel}(b) but apply for all panels.  A
Kendall-$\tau$ test (including limits) yields no statistical signature
for a correlation between {\HI} column density and $\Phi$
($0.4~\sigma$).  We do note that the $N({\HI})$ values appear to
increase toward the projected major axis ($\Phi=0^{\circ}$) and the
projected minor axis ($\Phi=90^{\circ}$).

To crudely estimate the degree to which the two largest $N({\HI})$
absorbers (which are most closely aligned with the projected axes) may
be outliers of the distribution of {\HI} absorbers, we calculated the
mean and standard deviation of the $N({\HI})$ excluding the two
largest $N({\HI})$ absorbers.  We obtained $\log \langle N({\HI})
\rangle = 14.47\pm 0.74$.  The two largest $N({\HI})$ absorbers lie at
$2.8~\sigma$ and $5.5~\sigma$ from $\langle N({\HI}) \rangle$.  For
this small sample, the $N({\HI})$ of the system most closely aligned
with the projected minor axis is an outlier of the $N({\HI})$
distribution in that it has a significantly larger column density.  A
statistical signature for an {\HI} column density enhancement toward
the projected major axis is less convincing.  

\begin{figure*}[th]
\epsscale{1.1}
\plottwo{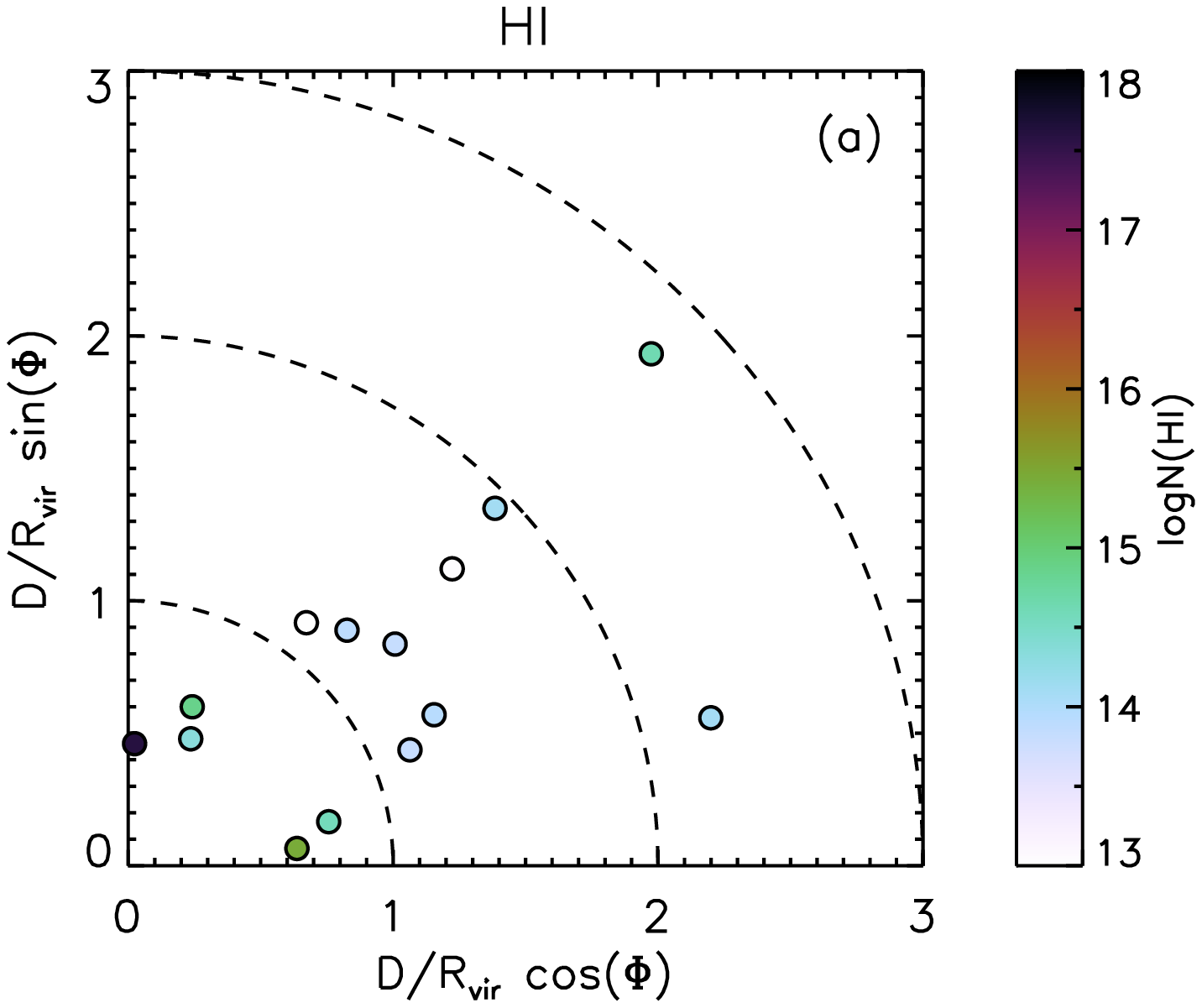}{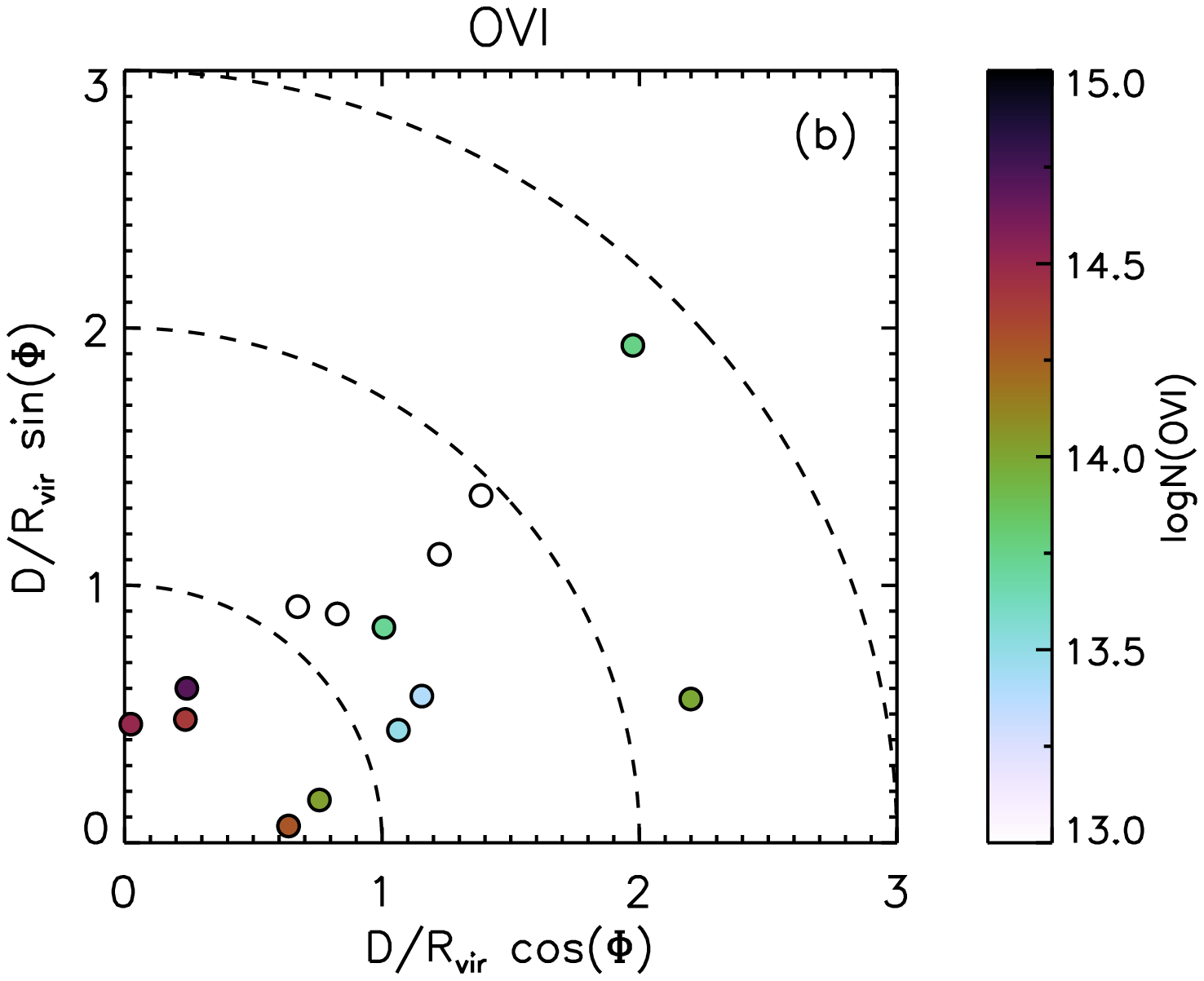}
\caption{Projected azimuthal and $D/R_{\rm vir}$ cloud locations for
  (a) {\HI} and (b) {\OVI}.  The horizontal axis represents the
  projected major axis and the vertical axis represents the projected
  minor axis.  The dashed lines represent curves of constant $D/R_{\rm
    vir}$.  Data points are colored according to the total system
  column density. Column density upper limits are plotted as open
  circles. }
\label{fig:XvsY}
\end{figure*}

We further examine the apparent trend for increasing $N({\HI})$ toward
the projected major and minor axes by symmetrically folding the
azimuthal angle about $45^{\circ}$, such that $0^{\circ}$ corresponds
to alignment along either the major or the minor projected axis and
$45^{\circ}$ corresponds to a $45^{\circ}$ azimuthal angle with
respect to either axis.  A Kendall-$\tau$ test (including limits) on
the folded distribution yields no statistical signature for an
anti-correlation between $N({\HI})$ and angular separation away from
the galaxy projected axes ($1.7~\sigma$).  If a trend exists, our
sample is too small to reveal a statistical significance.

In the case of {\OVI}, the azimuthal distribution of $N({\OVI})$
appears uniform, or flat, for all galaxy orientations.  Computing the
mean and standard deviation (omitting limits), we obtain $\log \langle
N({\OVI}) \rangle = 14.05 \pm 0.42$ and find no outlying
measurements with detected {\OVI} absorption.  

For the ratio $N({\OVI}) / N({\HI})$, the two absorbers within $\pm
10^{\circ}$ of the major and minor axes show a smaller ratio compared
to the data in the range $10^{\circ} \leq \Phi \leq
80^{\circ}$. Excluding these two absorbers and the two absorbers with
upper limits, the mean and standard deviation is $\log \langle
N({\OVI}) / N({\HI}) \rangle = -0.61 \pm 0.45$.  Compared to this
distribution, the $\Phi \simeq 6^{\circ}$ absorber is a $\sim\!
4~\sigma$ outlier and the $\Phi \simeq 87^{\circ}$ absorber is a
$\sim\!  8~\sigma$ outlier.  Since the $N({\OVI})$ distribution is
flat with azimuthal angle, whereas the $N({\HI})$ distribution
exhibits higher values near the projected axes, the smaller
$N({\OVI})/N({\HI})$ ratios in these two absorbers are driven by their
higher $N({\HI})$.  We infer that the chemical and/or ionization
conditions of the hot CGM are globally distributed, if patchy, for all
azimuthal angles more than $\pm 10^{\circ}$ away from the projected
major and minor axes (acknowledging some variation as suggested by the
few upper limits).
 
% ======== DISTANCE + ORIENTATION RESULTS ======== 
\subsection{Distance and Orientation}

For visualization purposes, in Figure~\ref{fig:XvsY}(a) and
Figure~\ref{fig:XvsY}(b), we illustrate the relationship between the
two-dimensional projected location of the absorbing gas and the system
total $N({\HI})$ and $N({\OVI})$, respectively.  The projected
geometric position of the quasar sightlines are computed with respect
to the virial radius using the relations $D/R_{\rm vir} \cos(\Phi)$
for projection along the galaxy major axis, and $D/R_{\rm vir}
\sin(\Phi)$ for projection along the galaxy minor axis. Data point
colors correspond to absorber column density according to the color
scale on the right. Upper limits are plotted as open circles.

Most galaxies probed at $D/R_{\rm vir} > 1$ in this sample have
moderate column density absorption with $\log N({\HI}) \simeq 14$
located at intermediate azimuthal angles, between $10^{\circ}$ and
$90^{\circ}$. In turn, the galaxies probed at $D/R_{\rm vir} < 1$
exhibit high column density absorption along their projected major and
minor axes. Though the sample is small and the realization of the data
may not be a full representation of a larger sample, our data suggest
a picture in which moderate column density {\HI} and {\OVI} gas is
distributed around galaxies out to greater than $2R_{\rm vir}$, with
higher column density {\HI} aligned near the galaxy projected major
and minor axes for $D < R_{\rm vir}$.

% =============== KINEMATICS AND VIRIAL MASS RESULTS ================
\section{Kinematics and Virial Mass}
\label{sec:kin}

\begin{figure*}[th]
\epsscale{1.0}
\plottwo{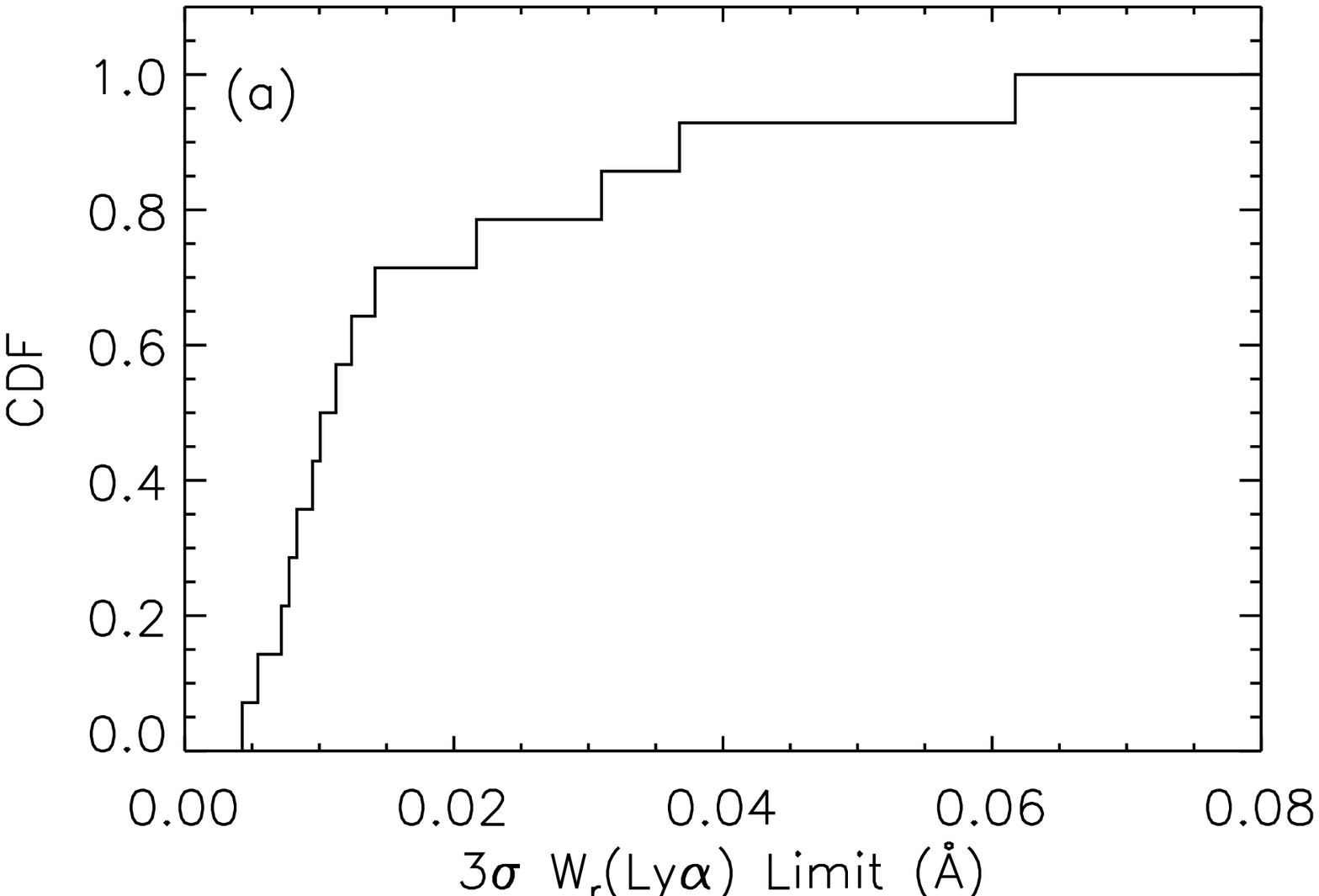}{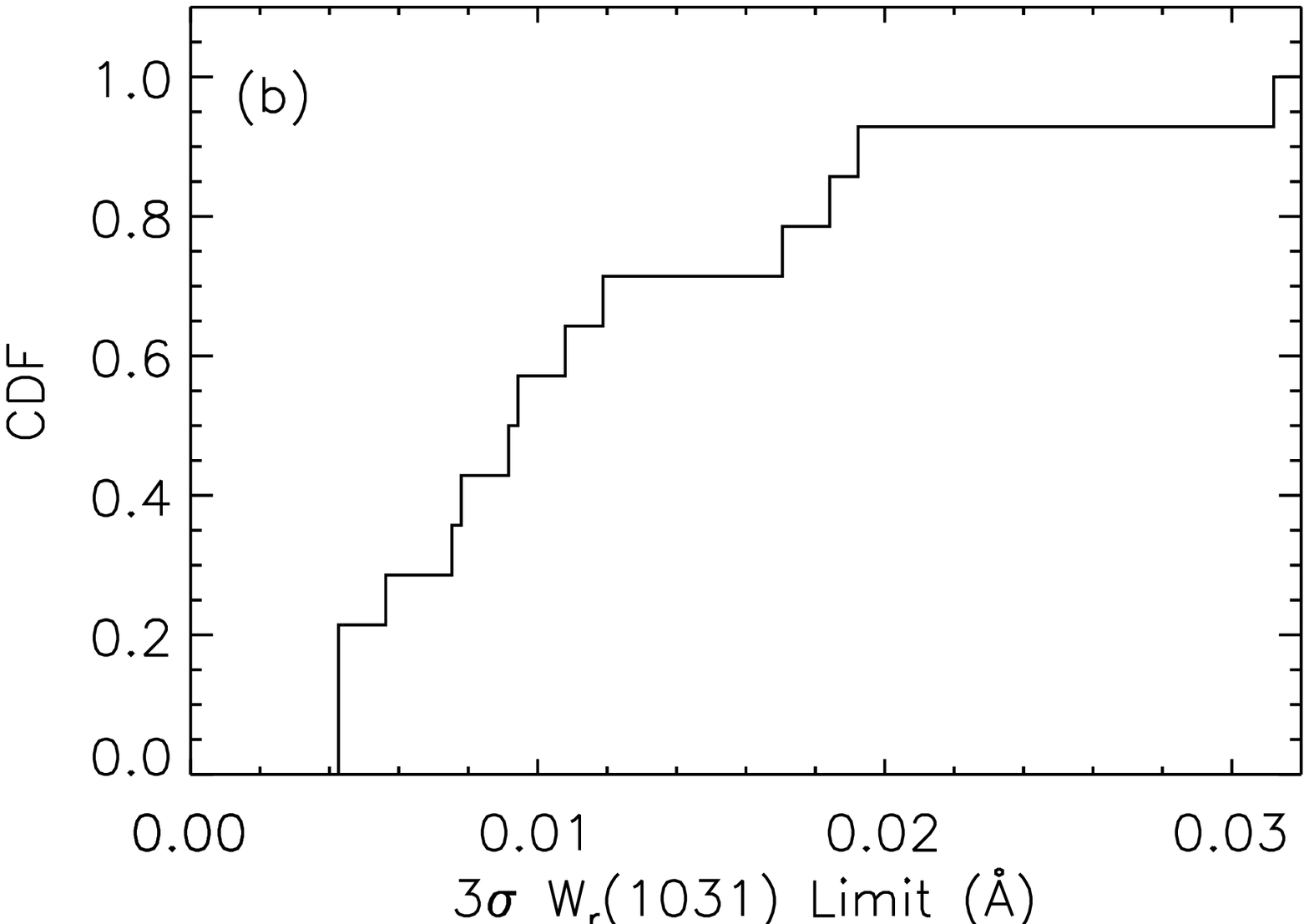}
\caption{Cumulative distribution function of the equivalent width
  detection limit in (a) the {\Lya} absorption, and (b) the
  {\OVIfirst} absorption.  Our sample has 100\% completeness in the
  detection sensitivity to $W_r({\Lya}) = 0.062$~{\AA} and $W_r(1031)
  = 0.032$~{\AA}. Only one {\Lya} absorption feature and one
  {\OVIfirst} absorption feature are measured with equivalent widths
  below the 100\% completeness sensitivity.  We have omitted those two
  features from our kinematic analysis.}
\label{fig:ABSCDF}
\end{figure*}

Characterizing the velocity distribution of the hot CGM is
instrumental for determining the physical origin and fate of both
{\HI} and {\OVI} absorbing gas in galactic environments.  If the
material is outflowing, velocity flows less than the galaxy halo
escape velocity might trace gas that is likely to recycle back into
the ISM and fuel star formation, whereas velocity flows greater than
the escape velocity might leave the CGM permanently and
chemically enrich the IGM.  If the material is infalling from the IGM
(or from satellite merging), the velocity distribution would provide
insights into mechanisms of how such gas mixes with the hot CGM or
eventually accretes into the ISM.

% ================ KINEMATIC COMPLETENESS RESULTS ======================
\subsection{Completeness}
\label{sec:completeness}

The detection threshold sensitivity for absorption is not uniform from
absorber to absorber due to varying signal-to-noise ratios, $S/N$, of
the quasar spectra.  Thus, for example, weak absorption at high
relative velocity that could be detected in a high $S/N$ spectral
region for one system, might not be detectable for a different system
appearing in a lower $S/N$ spectral region.  In conducting our
kinematic analysis, we first examine the non-uniformity of the
detection sensitivity.  

In Figure~\ref{fig:ABSCDF}, we present the cumulative distribution
(CDF) of the $3~\sigma$ equivalent width detection limits for both
{\Lya} and {\OVIfirst} absorption.  The detection thresholds are the
$3~\sigma$ equivalent width uncertainties for unresolved lines
\citep[cf.,][]{Schneider1993, Churchill2000} averaged over $\pm1000$
{\kms} relative to the galaxy redshift assuming unresolved absorption
lines.  The sample is 100\% complete for absorption features greater
than $W_r({\Lya}) = 0.062$~{\AA} and $W_r(1031) = 0.032$~{\AA},
indicating that we generally have higher $S/N$ spectral coverage for
{\OVI} absorption.  All but two of the individual absorption features
in the sample have measured equivalent widths above the 100\%
completeness level.  For our analysis, we removed these two features.

\begin{figure}[thb]
\epsscale{1.1}
\plotone{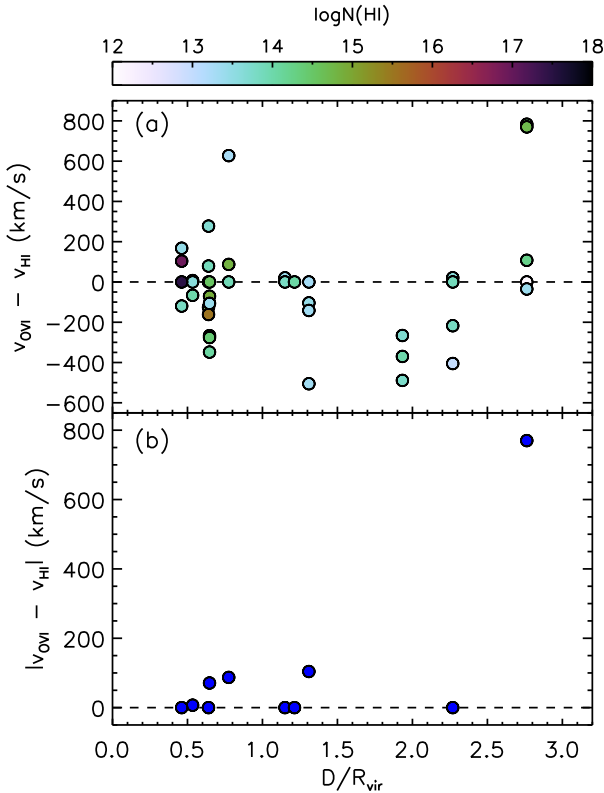}
\caption{(a) The velocity offsets of the {\HI} absorbing Voigt profile
  components ``clouds'' in a system with respect to the highest
  $N({\OVI})$ cloud in the system as a function of $D/R_{\rm vir}$.
  The colors of the data points are based upon the cloud $N({\HI})$ as
  given by the color bar legend.  (b) The velocity offset between the
  highest $N({\OVI})$ cloud and the highest $N({\HI})$ cloud in a
  system as a function of $D/R_{\rm vir}$.  In both panels, the dotted
  line at $v_{\hbox{\tiny OVI}} - v_{\hbox{\tiny HI}} = 0$ indicates
  no velocity offset.}
\label{fig:ComponentCompare}
\end{figure}

% ======== ALIGNMENT RESULTS ======== 
\subsection{Kinematic Alignment of\/ {\HI} and\/ {\OVI}}

\begin{figure*}[thb]
\epsscale{1.1}
\plottwo{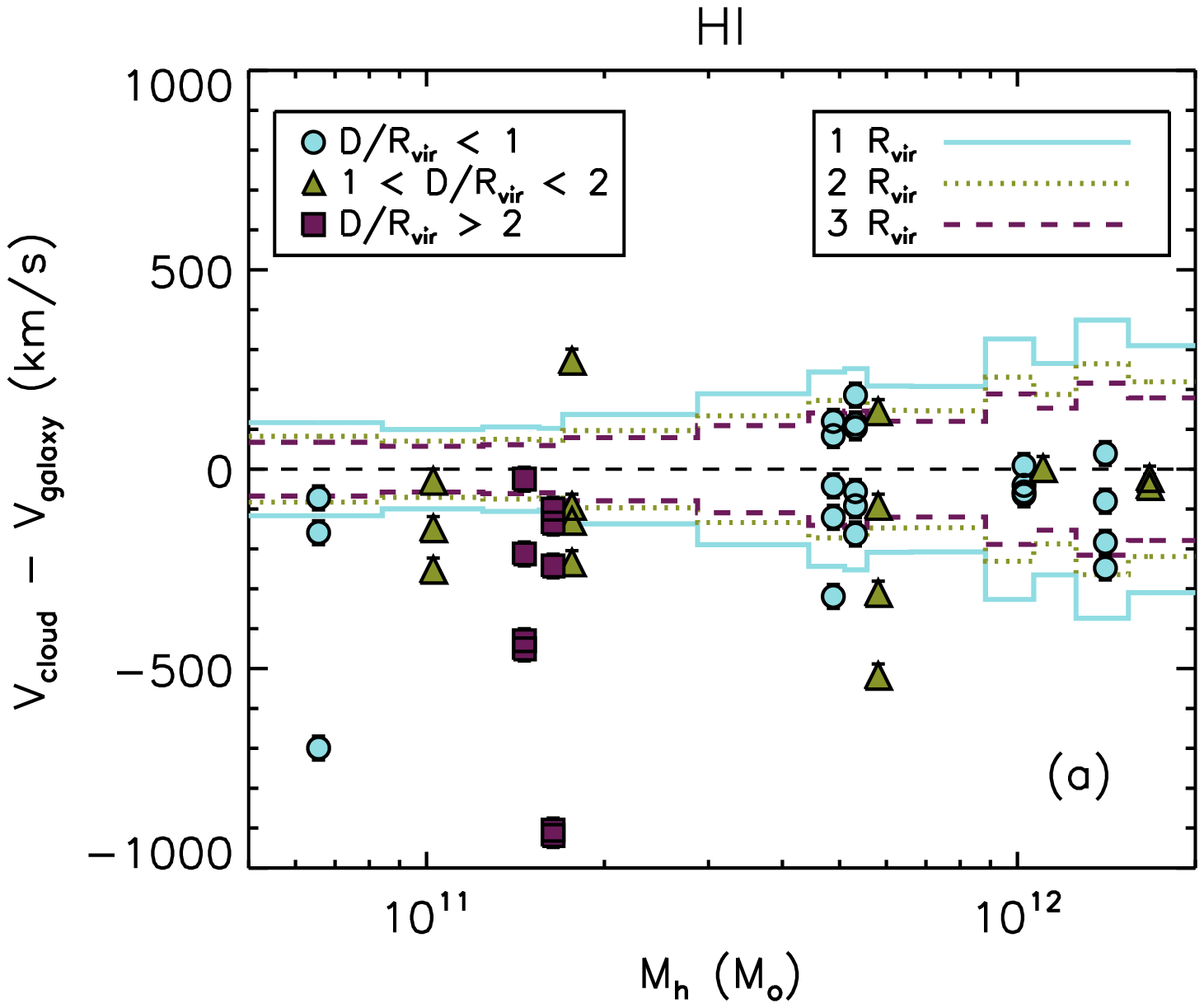}{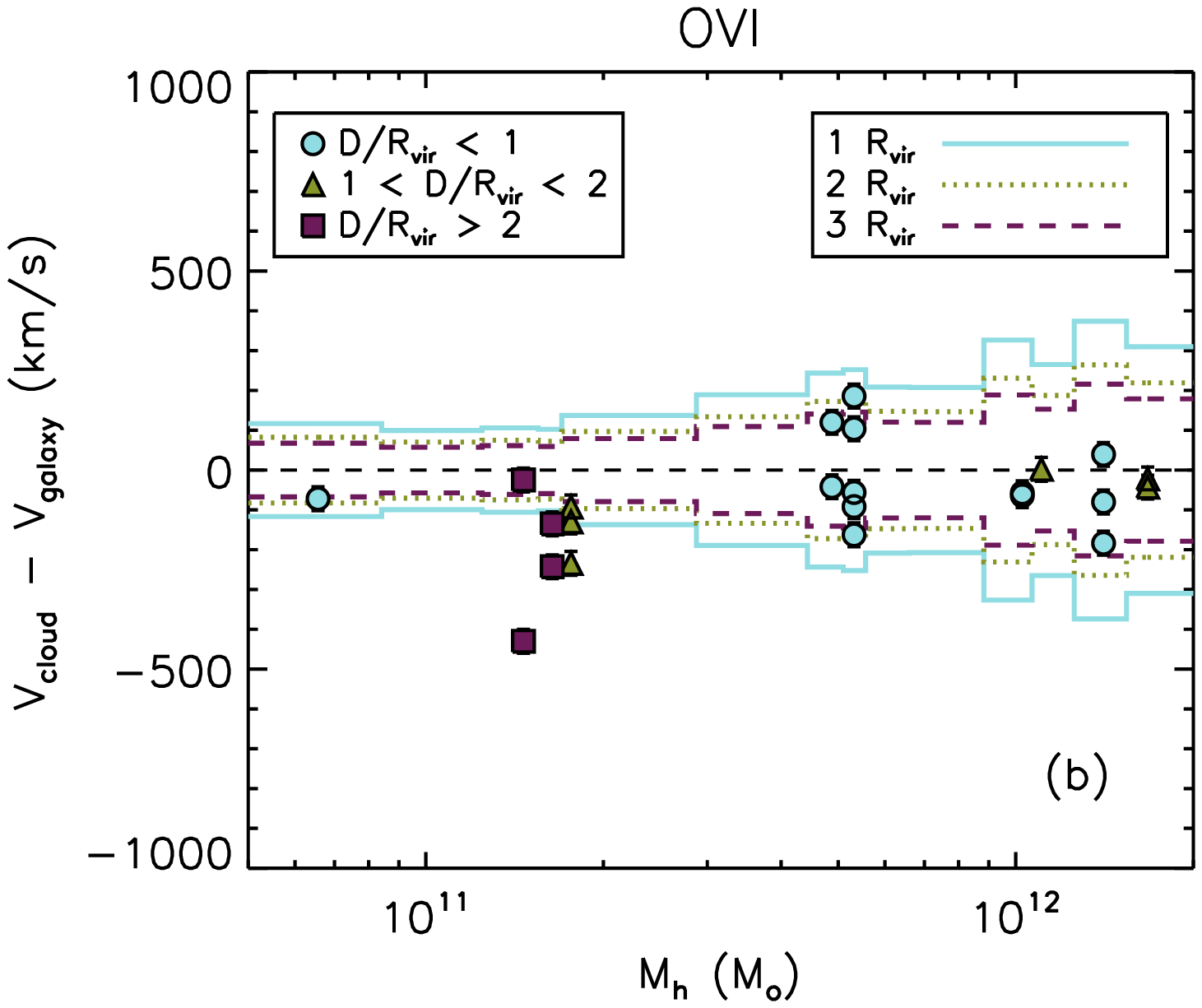}
\caption{Individual Voigt profile component (cloud) velocity offsets
  with respect to the galaxy systemic velocity as a function of virial
  mass, $M_{\rm\,h}$.  (a) {\HI} clouds.  (b) {\OVI} clouds.  Data are
  colored by $D/R_{\rm vir}$ bins. Vertical error bars are dominated
  by a $\simeq 30$~{\kms} uncertainty in the galaxy redshift.  The
  colored lines are the escape velocities, $v_{\rm esc}$, for each
  galaxy computed using Equation~\ref{eqn:vesc} for assumed cloud
  galactocentric distances $R=R_{\rm vir}$, $2R_{\rm vir}$, and
  $3R_{\rm vir}$.  At a given halo mass, points of a given color are
  to be compared to $v_{\rm esc}$ values having the same colored
  line.}
\label{fig:Vesc}
\end{figure*}

\begin{figure*}[bht]
\epsscale{1.1}
\plottwo{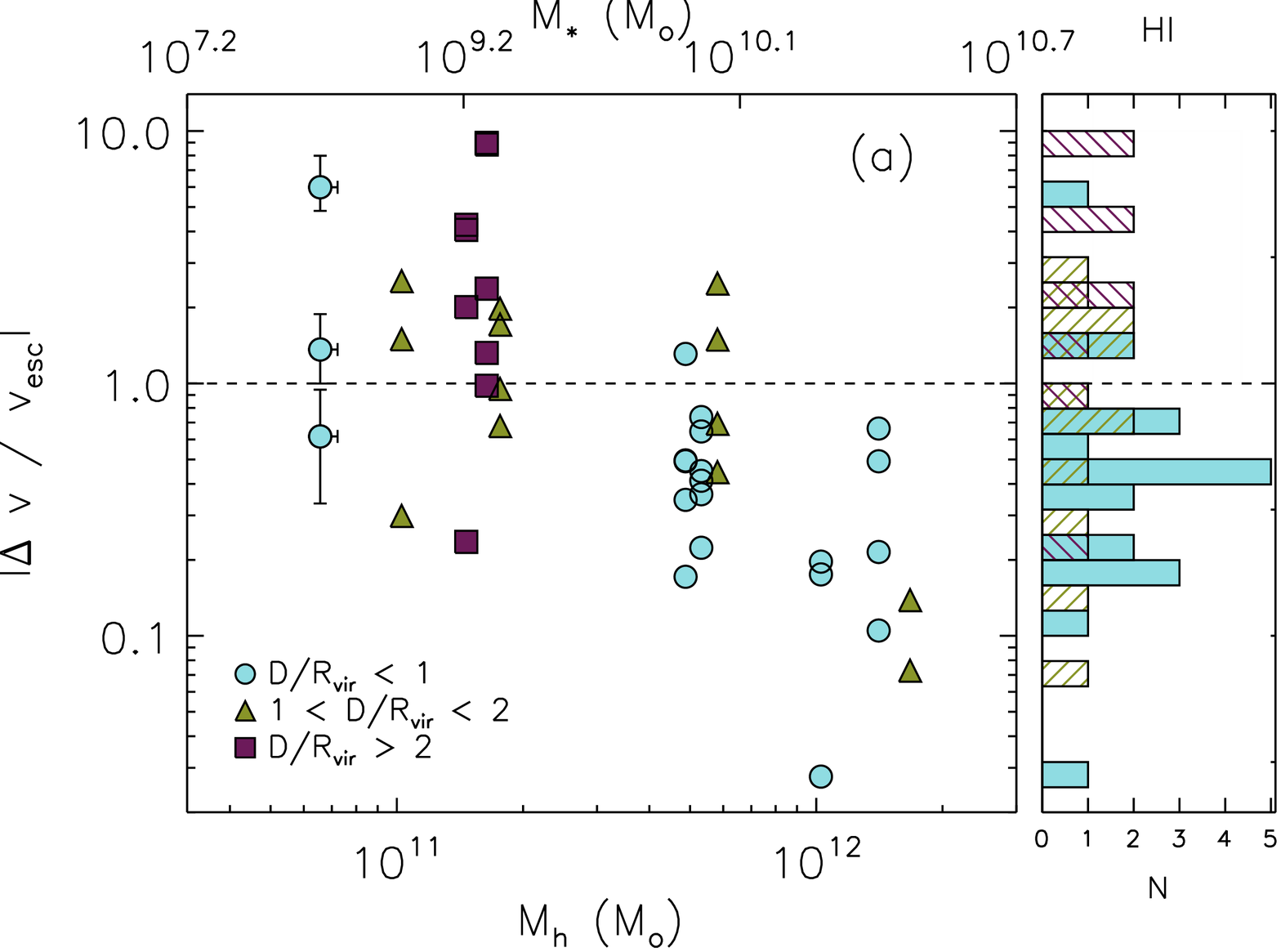}{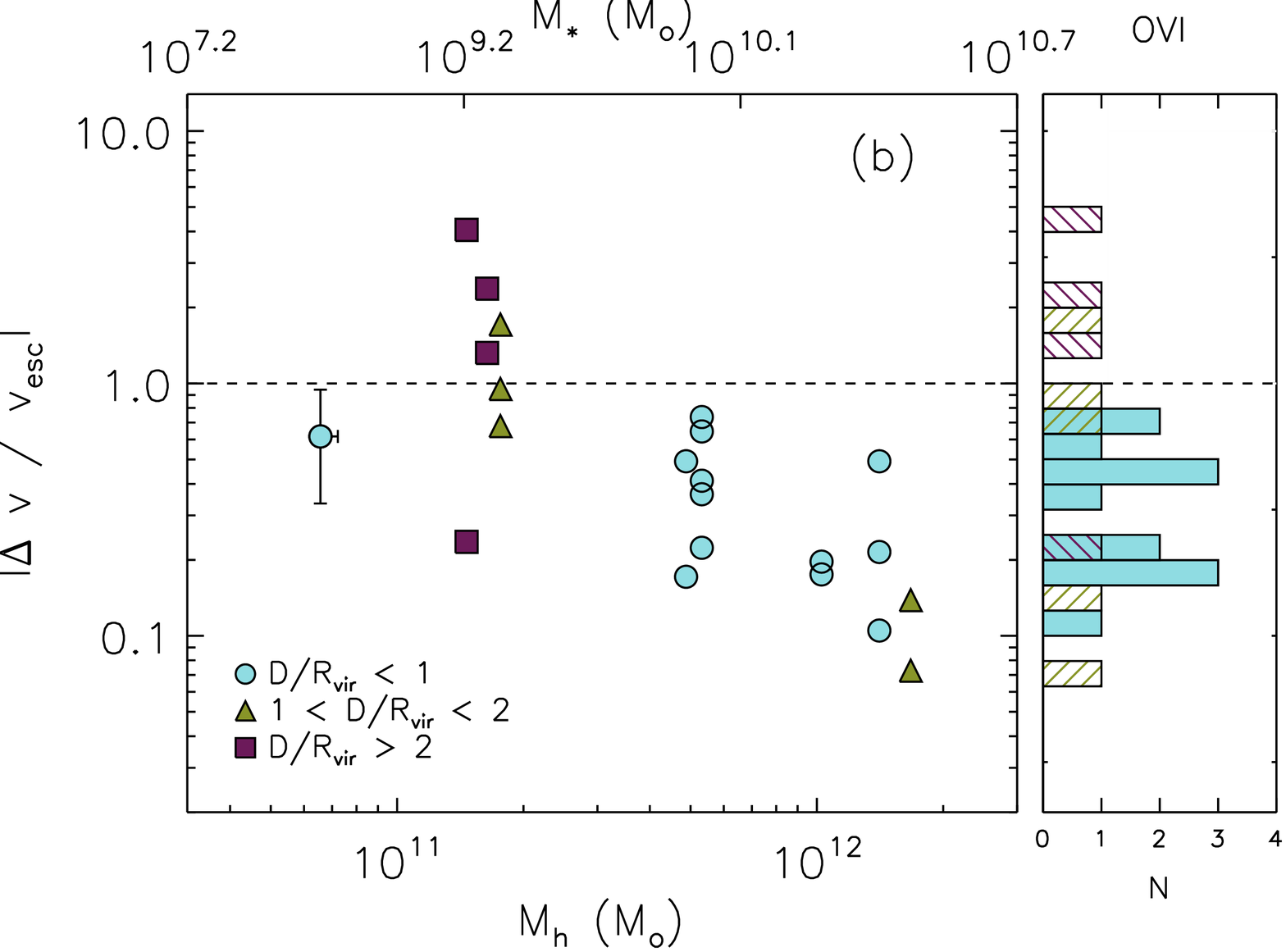}
\caption{The absolute relative velocity of the Voigt profile ``cloud''
  velocities with respect to the galaxy normalized to the escape
  velocity, $\left| \Delta v /v_{\rm esc}\right|$, as a function of
  virial and stellar mass, $M_{\rm\,h}$ and $M_{\ast}$.  (a) {\HI}
  clouds.  (b) {\OVI} clouds.  Data are colored by $D/R_{\rm vir}$
  bins using the same convention as for Figure~\ref{fig:Vesc}.  The
  escape velocity is computed assuming the clouds reside at
  galactocentric distance $R=D$, which yields lower limits on $\left|
  \Delta v /v_{\rm esc}\right|$.  Representative error bars for
  $\left| \Delta v /v_{\rm esc}\right|$ are provided for the lowest
  virial mass galaxy and account for the $\simeq 30$~{\kms}
  uncertainty in the galaxy redshift and in the virial mass propagated
  through Equation~\ref{eqn:vesc}.  The horizontal dashed line at
  $\left| \Delta v /v_{\rm esc}\right|=1$ represents cloud offset
  velocities equal to the halo escape velocity.  Histograms (right
  subpanels) provide the number of clouds in equal logarithmic bins
  for all virial masses. The fraction of clouds with velocity offsets
  greater than the halo escape velocity increases with decreasing
  stellar and halo mass.}
\label{fig:MassEscape}
\end{figure*}

In Figure~\ref{fig:ComponentCompare}(a), we plot the velocity offset
between the Voigt profile component ``clouds'' with the highest
$N({\OVI})$ and each {\HI} absorbing cloud as a function of $D/R_{\rm
  vir}$.  The dotted line at $v_{\hbox{\tiny OVI}} - v_{\hbox{\tiny
    HI}} = 0$ indicates no velocity offset between the {\HI} clouds
and the highest $N({\OVI})$ cloud.  Most {\HI} and {\OVI} absorbing
clouds are clustered within $\sim\!  500$~{\kms}.

The data also reveal that the highest $N({\OVI})$ cloud in a system
does not align kinematically with the highest $N({\HI})$ cloud in
every case.  To further illustrate, we plot the velocity offset
between the highest $N({\OVI})$ cloud and the highest $N({\HI})$ cloud
in a system in Figure~\ref{fig:ComponentCompare}(b). In 5 of 10
systems with detected {\OVI} absorption, we observe a velocity offset
between the bulk of the neutral hydrogen and the bulk of the {\OVI}.
In three of these cases, the velocity offset is $\sim\!100$~{\kms}.
Since a velocity offset implies spatially separated absorbing clouds,
we can infer physically distinct phases of gas (different densities,
temperatures, and metallicities) in roughly half of the sightlines
through the CGM, as probed using {\OVI} and {\HI} as a tracer of the
gas phase. Examining absorption from the low ions such as {\CII},
{\SiII}, etc., would be instrumental in examining whether this is the
case.  Unfortunately, the mean $N({\HI})$ for our sample is roughly
1.5 dex below the threshold where low ion metals can be detected in
spectra with moderate signal-to-noise ratios \citep{Hellsten1997}.

% ============= CLOUD KINEMATICS RESULTS ===========
\subsection{Kinematics and Escape Velocity}
\label{sec:cloudkin}

In Figure~\ref{fig:Vesc}, we plot the velocity difference between the
individual Voigt profile component ``clouds'' and the galaxy systemic
velocity, $\Delta v = v_{\rm cld} - v_{\rm gal} = c(z_{\rm cld} -
z_{\rm gal})/(1+z_{\rm gal})$, as a function of halo mass,
$M_{\rm\,h}$.  Data point colors denote the $D/R_{\rm vir}$ location
of the absorption, where sky-blue corresponds to $D/R_{\rm vir} \le
1$, green between $1 < D/R_{\rm vir} \le 2$, and $D/R_{\rm vir} > 2$
is magenta. For our sample, we find that $\sim\!70$\% of {\HI}
absorbing components lie within $\pm200$~{\kms} of the galaxy systemic
velocity, $\sim\!80$\% lie within $\pm300$~{\kms}, and $\sim\!90$\%
lie within $\pm500$~{\kms}.  For {\OVI}, we find $\sim\!90$\%,
$\sim\!95$\%, and $100$\% lie within $\pm200$, $\pm300$, and
$\pm500$~{\kms}, respectively.

To investigate whether clouds have relative line of sight velocities
that exceed or do not exceed the escape velocity of the halo in which
they reside, we computed the escape velocity for each galaxy
\citep[cf.,][]{Steidel2010},
\begin{equation}
v_{\rm esc}^{\,2}(R\,)\,=\,\frac{2GM_{\rm\,h}}{R} 
\frac{\ln[1+ c(R/R_{\rm vir})]}
     {\ln(1+c)-c/(1+c)} \, ,
\label{eqn:vesc}
\end{equation}
for each galaxy at $R=R_{\rm vir}$, $2R_{\rm vir}$, and $3R_{\rm
  vir}$, where $R$ is galactocentric distance, and superimposed the
results on Figure~\ref{fig:Vesc} as colored lines corresponding to the
three values of $R$.  A \cite{Navarro1997} (NFW) dark matter halo
profile is assumed with mean concentration parameter,
$c\,(M_{\rm h}, z_{\rm gal})$, computed from the relation of
\cite{Bullock2001}.  The curves are not smooth with increasing virial
mass due to the redshift dependence of the concentration parameter.

The galactocentric distances of the clouds are not known, only
constrained to lie at $R\geq D$.  Assuming cloud galactocentric
distances at multiples of the virial radius, and comparing
same-colored points and curves, we find $\sim\!40$\% of {\HI} cloud
components have relative velocities in excess of the galaxy escape
velocity.  Roughly $60$\% of the clouds that reside outside the virial
radius (green and magenta data) have velocities exceeding the escape
velocity, whereas only $\sim\!15$\% of clouds, if they reside at the
virial radius (sky-blue data/lines), have greater than escape velocities.

% =============== GALAXY ESCAPE VELOCITY RESULTS =============
\subsection{Differential Kinematics}
\label{sec:vesc}

In Figure~\ref{fig:MassEscape}, we show the absolute relative velocity
of the absorption with respect to the galaxy normalized to the escape
velocity, $\left| \Delta v /v_{\rm esc}\right|$, versus virial mass.
We also show the galaxy stellar mass, $M_{\ast}$, based upon the
stellar mass to halo mass functions of \citet{Moster2010}.  Data
points are colored by $D/R_{\rm vir}$ using the same designations as
in Figure~\ref{fig:Vesc}.  For this exercise, we assume $R=D$ for the
absorbing clouds, which provides the upper limit to the escape
velocity.  Thus, the points plotted in Figure~\ref{fig:MassEscape} are
lower limits on $\left| \Delta v /v_{\rm esc}\right|$.  Characteristic
error bars are shown on the left-most data points in each panel.
Points that lie above the dotted line at $\left| \Delta v /v_{\rm
  esc}\right|=1$ are clouds that have velocities in excess of the
galaxy escape velocity and, if outflowing, could be unbound. The
histogram on the right shows the total number of clouds in $\left|
\Delta v /v_{\rm esc}\right|$ bins for all halo masses.

For {\HI} absorption, we computed the fraction of clouds with $\left|
  \Delta v /v_{\rm esc}\right| \leq 1$, i.e., those that can be
inferred to be gravitationally bound to their host halo.  We divide
the sample of clouds into several subsamples based upon $D/R_{\rm
  vir}$ and virial mass, $M_{\rm\,h}$.  The $D/R_{\rm vir}$ ranges are
$(0,1]$, $(1,2]$, $(2,3]$, and $[0.3)$ for each mass range,
$M_{\rm\,h} < 10^{11.5}$~M$_{\odot}$, $M_{\rm\,h} >
10^{11.5}$~M$_{\odot}$, and ``all'' $M_{\rm\,h}$, where $M_{\rm\,h} =
10^{11.5}$~M$_{\odot}$ is the median virial mass of the sample.
Comparing the ``bound fraction'' in each of these subsamples allows a
differential characterization of {\HI} kinematics.

%As seen in Figure 9, there is a natural division in the mass range,
%seen as a clear gap in Mh.

\begin{deluxetable}{lccc}
\tablecolumns{4}
\tablewidth{0pt}
\setlength{\tabcolsep}{0.06in}
\tablecaption{Bound Fractions of {\HI} Clouds\label{tab:bound}}
\tablehead{
  \colhead{(1)} &
  \colhead{(2)} &  
  \colhead{(3)} &  
  \colhead{(4)} \\[2pt]  
  \colhead{$D/R_{\rm vir}$}  &
  \colhead{all $M_{\rm\,h}$} &
  \colhead{$M_{\rm\,h} > 10^{11.5}$} & 
  \colhead{$M_{\rm\,h} < 10^{11.5}$} \\[1pt] 
  \colhead{range} &
  \colhead{[M$_{\odot}$]} &
  \colhead{[M$_{\odot}$]} &
  \colhead{[M$_{\odot}$]} }
\startdata
$0< D/R_{\rm vir} \leq 1$ & $0.85_{-0.13}^{+0.08}$ & $0.94_{-0.12}^{+0.05}$ & $0.33_{-0.28}^{+0.41}$ \\[3pt]
$1< D/R_{\rm vir} \leq 2$ & $0.54_{-0.17}^{+0.17}$ & $0.67_{-0.28}^{+0.21}$ & $0.43_{-0.22}^{+0.25}$ \\[3pt]
$2< D/R_{\rm vir} \leq 3$ & $0.11_{-0.09}^{+0.21}$ & $\cdots$            & $0.11_{-0.09}^{+0.21}$ \\[3pt]
$0< D/R_{\rm vir} \leq 3$ & $0.59_{-0.09}^{+0.09}$ & $0.87_{-0.11}^{+0.07}$ & $0.26_{-0.11}^{+0.14}$ \\[3pt]
\\[-18pt]
\enddata
\end{deluxetable}

The {\HI} cloud bound fractions are presented in
Table~\ref{tab:bound}.  Column (1) lists the $D/R_{\rm vir}$ range,
and columns (2), (3), and (4) list the bound fractions for the three
mass ranges.  The bound fractions are $n_1/(n_1+n_2)$, where $n_1$ is
the number of clouds with $\left| \Delta v /v_{\rm esc}\right| \leq 1$
and $n_2$ is the number of clouds with $\left| \Delta v /v_{\rm
    esc}\right| > 1$.  The quoted uncertainties assume a binomial
distribution \citep[see][]{Gehrels1986} and were computed using
incomplete $\beta$ functions for a confidence level of 84.13\% (single
sided $1~\sigma$).  We remind the reader that we measure line of sight
velocities. Despite having constraints on the galaxy inclinations and
azimuthal angles on the sky, deprojecting the gas velocities is an
intractable problem due to significant uncertainty in the true gas
motions which are a result of the complex interplay between outflow
geometry, environmental conditions, and inflow dynamics.

Examining column (2) of Table~\ref{tab:bound}, we find that the bound
fraction decreases as $D/R_{\rm vir}$ increases.  In other words, the
proportion of clouds with greater than escape velocities increases
with increasing projected distance relative to the virial radius.  On
average, $\sim\!40$\% of the clouds could be inferred to be escaping
the halo for $D/R_{\rm vir} \leq 3$ for all virial
masses represented in the sample, assuming the clouds are outflows
(we reserve further discussion on this point until
Section~\ref{sec:inorout}).

Comparing columns (3) and (4) of Table~\ref{tab:bound}, we find that,
in each and every $D/R_{\rm vir}$ range, higher mass halos have a
larger fraction of bound clouds than do lower mass halos.  Or,
alternatively, the fraction of clouds with $\left| v \, \right| >
v_{\rm esc}$ is larger in lower mass halos than in higher mass halos.
On average, $\sim\!75$\% of the clouds have $\left| v \, \right| >
v_{\rm esc}$ for lower mass halos, whereas only $\sim\!10$\% of the
clouds have $\left| v \, \right| > v_{\rm esc}$ for higher mass halos.

When interpreting the trends presented in Table ~\ref{tab:bound}, we
must be careful to consider the selection effect that, in small
samples characterized by a pre-selected $D$ range, lower mass halos
are preferentially probed at larger $D/R_{\rm vir}$ because they have
smaller virial radii than larger mass halos.  Thus, the majority of
the absorbers in the $M_{\rm\,h} < 10^{11.5}$~M$_{\odot}$ subsample
are probed at $D/R_{\rm vir} > 1$.  The trend that the bound fraction
decreases as $D/R_{\rm vir}$ increases may be enhanced in our sample
due to the increase in the relative number of lower mass galaxies at
larger $D/R_{\rm vir}$.  This is corroborated by the result that
higher mass halos have a larger fraction of bound clouds than do lower
mass halos in each finite $D/R_{\rm vir}$ range, which likely does not
suffer from any selection bias and is a more robust finding.

Examining {\OVI} absorbers, we find associated {\OVI} absorption in
only $\sim\!40$\% of the {\HI} clouds in and around lower mass halos
as compared to $\sim\!85$\% around higher mass halos. Given the flat
{\HI} column density distribution for our sample, the lower number of
detected {\OVI} clouds in lower mass halos suggests conditions favoring
higher {\OVI} column densities are less common out to $D \simeq
300$~kpc of lower mass halos than for higher mass halos.

For {\OVI} absorbers, as shown in Figure~\ref{fig:MassEscape}(b), the
clouds have a bound fraction of $0.82_{-0.12}^{+0.09}$ for all halo
masses in the sample.  For higher mass halos, the bound fraction is
$1.00_{-0.12}^{+0.00}$ and for lower mass halos the bound fraction is
$0.50_{-0.22}^{+0.22}$.  We thus can infer that {\OVI} absorbing gas
is more common in higher mass halos and is primarily bound to the
halo, whereas {\OVI} absorbing gas is less commonly found in the
vicinity of lower mass halos and only half of the {\OVI} absorbing
clouds are bound.

\section{Discussion}
\label{sec:discussion}

% ======= CGM EXTENT AND GEOMETRY DISCUSSION ===========
% ======== DISTANCE DISCUSSION ======== 
\subsection{Spatial Extent of the Hot CGM}
\label{sec:distancediscussion}

Our small sample of 14 galaxy-absorber pairs is similar to that of the
COS-Halos project \citep{Tumlinson2013}, but with a few differences.
First, our sample probes {\HI} and {\OVI} absorption out to 300~kpc,
whereas the COS-Halos sample probes out to 150~kpc. Second, our sample
covers a slightly smaller range of virial mass, $10.8 \le
\log(M_{\rm\,h}/M_{\odot}) \le 12.2$, as compared to COS-Halos, $11.3
\le \log(M_{\rm\,h}/M_{\odot}) \le 13.3$.  We thus probe to 0.5 dex
lower in virial mass, but do not probe the highest full decade of the
COS-Halos sample.  The redshift coverage is roughly identical to that
of COS-Halos ($z \la 0.3$), but for the single galaxy in our sample at
$z=0.66$.

We note that the selection criteria between the two surveys are not
too dissimilar, since we also selected our galaxies with no {\it a
  priori\/} pre-disposition to {\HI} absorption in the background
quasar spectra.  We do not have the data to estimate the specific star
formation rates of the galaxies in our sample.  However, since
\citet{Tumlinson2013} conclude there is very weak evidence for a
difference in the detection frequency of {\HI} between
``star-forming'' and ``passive'' galaxies, there should be
little-to-no ambiguity in comparing the neutral hydrogen between
samples.

To a $3~\sigma$ equivalent width detection sensitivity of $W_r({\Lya})
= 0.05$~{\AA} (100\% completeness, corresponding to $\log N({\HI}) =
13$ for $b=30$~{\kms}), we find {\HI} absorption is present out to
300~kpc for 13 of 14 galaxies in our sample, indicating that {\HI} gas
is clearly present out to $\sim\! 3$ projected virial radii.  For
$D/R_{\rm vir}< 1$, we measure a mean system total column density of
$\log \langle N({\HI}) \rangle = 15.9\pm 1.6$, which is in good
agreement with the column densities found in this region for the
COS-Halos sample, which probes out to $D/R_{\rm vir} \simeq 0.75$.
For $D/R_{\rm vir} > 1$, we find $\log \langle N({\HI}) \rangle =
14.1\pm 0.5$, a value $\simeq 1.8$ dex lower.  This behavior would
suggest a transition at $D/R_{\rm vir} \simeq 1$ in the physical
nature of {\HI}, possibly due to a changes in cloud densities, sizes,
or ionization conditions.

The behavior of {\HI} that we described above is consistent with the
conclusion drawn from \citet{Tumlinson2013}, that there is significant
``evolution'' in the {\HI} properties between the regions $D <
200$~kpc and outside this region, as based upon their comparison with
several other studies of {\HI} absorption around galaxies (see their
Section 5.1).  In our smaller sample representing the lower mass range
of COS-Halos, this ``evolution'', or transition, appears to set in at
$D\simeq 100$~kpc.  This could imply that lower mass halos have
smaller physical extent than higher mass halos \citep[also
  see][]{MAGIICAT3, Ford2013mass}. Consistent with the conclusions of
\citet{Tumlinson2013} and \citet{Stocke2013}, we also infer that,
based upon the behavior of {\HI} absorption, the virial radius appears
to be a transition region between the CGM and the IGM.  Our data
suggest this region is quite extended and is not abrupt.  However,
this does not preclude that possibility that metals from the ISM are
being transported through the CGM to several virial radii and out to
the IGM.

To a $3~\sigma$ detection sensitivity of $W_r(1031) = 0.032$~{\AA}
(100\% completeness, corresponding to $\log N({\OVI}) = 13.5$ for
$b=20$~{\kms}), we find {\OVI} absorption is present out to the impact
parameter limit of our survey ($\sim\!300$~kpc) for 11 of 14 galaxies
in our sample.  This corresponds to {\OVI} absorption as far out as
$D/R_{\rm vir} \simeq 2.7$, which could suggest that {\OVI} enrichment
for the host galaxy as far as $\sim\!  3$ virial radii.  We do not
find an anti-correlation for $N({\OVI})$ with $D$ nor with $D/R_{\rm
  vir}$, though the five detections at $D/R_{\rm vir} < 1$ are the
five highest $N({\OVI})$ absorbers.

Out to their survey limit of $D=150$~kpc ($D/R_{\rm vir} \simeq
0.75$), COS-Halos finds that {\OVI} absorption with $\log N({\OVI})
\geq 14.2$ is preferentially found in the CGM of star forming galaxies
\citep{Tumlinson2011}.  As stated above, we cannot address a
comparison between {\OVI} absorption and the star forming properties
of the galaxies in our smaller sample.  However, it would be
interesting to do so given our deeper detection sensitivity to {\OVI}
absorption.  COS-Halos is roughly 20\% complete (9/42) for detections
below $\log N({\OVI}) = 14.0$ and 100\% complete to $\log N({\OVI}) =
14.2$, whereas we are 100\% complete to $\log N({\OVI}) = 13.5$.  Of
interest is that the ``passive'' galaxies are among the most massive
galaxies in the COS-Halos sample and lie in the mass range not
represented in our sample.  If we speculate that this implies our
lower mass galaxies are drawn from the same population as the
star-forming population represented in the COS-Halos survey, then we
have found examples where the CGM of star forming galaxies can have
detectable absorption weaker than $\log N({\OVI}) = 14.2$.

We find that the distribution of the total system ratio $N({\OVI}) /
N({\HI})$ is consistent with being flat out to $D= 290$~kpc and
$D/R_{\rm vir} \simeq 2.7$.  However, in terms of individual
clouds (VP components), there is a higher incidence of {\OVI}
absorption in {\HI} clouds for higher mass halos than in lower mass
halos.  Dividing the sample by the median virial mass of $\log
M_{\rm\,h}/M_{\odot} =11.5$, {\OVI} absorption is found in only
$\sim\! 40$\% of the {\HI} clouds in the around lower mass halos as
compared to $\sim\! 85$\% around higher mass halos. Since the system
total $N(\HI)$ is fairly flat, the smaller fraction of detected {\OVI}
clouds in lower mass halos suggest conditions favoring {\OVI} are less
common out to $D\simeq 300$ kpc of lower mass halos than for higher
mass halos.

\subsection{Geometric Distribution of the Hot CGM}
\label{sec:orientationdiscussion}

% orientation
Using {\MgII} absorbers, \cite{Kacprzak2012-PA} and \cite{Bouche2012}
have shown cool/warm CGM gas is more frequently found to be aligned
with either the galaxy projected major or minor
axes. \cite{Bordoloi2011} finds stronger {\MgII} is preferentially
aligned with the projected minor axis.  \cite{Kacprzak2011} find that
{\MgII} equivalent widths correlate with galaxy inclination when
scaled by the impact parameter.  Based upon these results, the authors
have suggested wind driven material may be responsible for the
enhanced absorption strengths aligned with the galaxy minor axes,
whereas the inclination correlation may indicate a planer distribution
(also seen as a projected major axis alignment) of absorbing gas,
perhaps accreting from the IGM.

For our sample, our data suggest that higher {\HI} column density gas
is preferentially found within $\pm 10^{\circ}$ of the major and minor
axes (inside the projected virial radius).  However, this is not a
statistically significant result. 

In the case of {\OVI}, as seen in Figure~\ref{fig:PAPanel}(b), the azimuthal
distribution of $N({\OVI})$ is statistically consistent with being
flat.  However, the ratio $N({\OVI}) / N({\HI})$ for the two absorbers
within $\pm 10^{\circ}$ of the major and minor axes have the smallest
values and are statistical outliers (to better than $4~\sigma$) as
compared to the values in the range $10^{\circ} \leq \Phi \leq
80^{\circ}$.  This would suggest that, on average, the chemical and/or
ionization conditions of the hot CGM are fairly uniform in their
geometrical distribution around galaxies for azimuthal angles more
than $\pm 10^{\circ}$ away from the projected major and minor axes
(acknowledging some variation as suggested by the fact that not all
{\HI} clouds exhibit {\OVI} absorption).

The higher {\HI} column density gas clouds with lower $N({\OVI}) /
N({\HI})$ ratios, but with typical $N({\OVI})$, that reside within
$\pm 10^{\circ}$ of the major and minor axes, may reflect the presence
of multi-phase gas at these geometric projections, with most of the
{\HI} associated with higher density, lower ionization gas.  The
ionization conditions, densities, and temperatures of the $\Phi \simeq
87^{\circ}$ (minor axis aligned) absorber (Q1317$+$2743 at $z=0.6605$)
are clearly multi-phase in nature \citep{Kacprzak2012-1317}.  No
detailed study has been conducted on the $\Phi \simeq 6^{\circ}$
(major axis aligned) absorber (Q1136$-$1334 at $z=0.2044$).  However,
the COS/G130M spectrum covers several low and intermediate ionic
transitions, including but not limited to {\CII}~$\lambda 1035$,
{\CIII}~$\lambda 977$, {\SiII}~$\lambda\lambda 1190,1193$, and
{\SiIII}~$\lambda 1206$.  Of these, {\CIII} and {\SiIII} are clearly
detected in absorption, {\CII} may be weakly detected, and {\SiII}
resides in a very noisy region of the spectrum.

We might infer that the two absorbers within $\pm 10^{\circ}$ of the
major and minor axes have larger $N({\HI})$ columns because they are
multi-phase systems, whereas the systems at greater angular separation
from the galaxy projected axes trace the hot CGM.  This would be
consistent with the findings of \citet{Kacprzak2012-PA}, who report an
increase in the frequency of {\MgII} absorbers with azimuthal
locations aligned with the projected major and minor axes.
Unfortunately, the $N({\HI})$ values for the remaining absorbers in
our sample are roughly 1.5 dex below the threshold where low ion
metals can be detected, even for solar metallicity gas
\citep{Hellsten1997}.

\subsection{Interpreting the Kinematics}
\label{sec:inorout}

\begin{figure*}[bht]
\epsscale{1.1}
\plotone{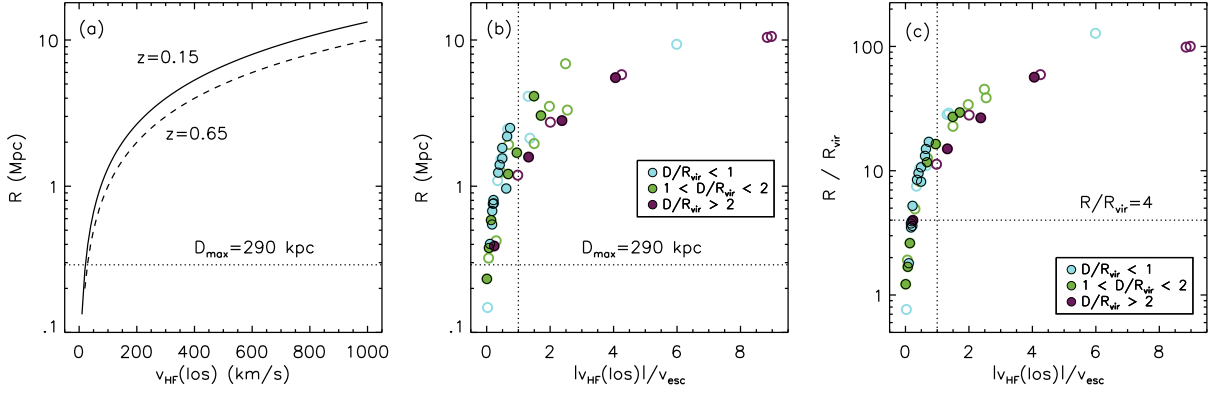}
\caption{The proper galactocentric distances, $R$, to the absorbing
  clouds assuming that the observed cloud-galaxy velocity offsets are
  pure line of sight Hubble flow, i.e., $\left| \Delta v \, \right| =
  \left| v_{\hbox{\tiny HF}}({\rm los})\right|$.  (a) $R$ versus
  $v_{\hbox{\tiny HF}}({\rm los})$ for $z_{\rm gal}=0.15$ and $z_{\rm
    gal} = 0.65$, illustrating Eq.~\ref{eq:hflow}.  At each $z_{\rm
    gal}$, the curves are independent of $D$ for $R \gg D$.
  The horizontal line at $R = D_{\rm max} = 290$~kpc indicates the
  maximum impact parameter of the sample.  (b) $R$ as a function of
  $\left| v_{\hbox{\tiny HF}}({\rm los})\right| /v_{\rm esc}$ for the
  individual clouds in our sample. Points are colored by their
  $D/R_{\rm vir}$ location using the color scheme employed for
  Figures~\ref{fig:Vesc} and \ref{fig:MassEscape}.  Filled points are
  clouds for which {\OVI} absorption is detected with the {\HI}
  absorption and open points are {\HI} only clouds.  The vertical line
  is $\left| v_{\hbox{\tiny HF}}({\rm los})\right| /v_{\rm esc} =
  1$. (c) $R/R_{\rm vir}$ as a function of $\left| v_{\hbox{\tiny
      HF}}({\rm los})\right| /v_{\rm esc}$.  The vertical line is
  $\left| v_{\hbox{\tiny HF}}({\rm los})\right| /v_{\rm esc} = 1$ and
  the horizontal line is $R/R_{\rm vir}=4$, the spatial location
  around a halo where Hubble flow dominates over dark matter
  accretion.}
\label{fig:hflow}
\end{figure*}

Constraining the galactocentric distances of the absorbing clouds is
central to interpreting the differential behavior in the bound
fraction of {\HI} clouds (see Figure~\ref{fig:MassEscape} and
Table~\ref{tab:bound}).  The data place only lower limits of $R \geq
D$ on their galactocentric distance and $R/R_{\rm vir} \geq D/R_{\rm
  vir}$ on their distances relative to the virial radius; we
cannot definitively determine whether a given absorber arises in the
IGM beyond $R=4R_{\rm vir}$, where Hubble flow begins to dominate the
peculiar velocities \citep[see][]{Cuesta2008}.

Based upon SPH simulations, \citet{Oppenheimer2009} argue that weaker
{\OVI} absorbers, $\log N({\OVI}) \simeq 14$, may trace the old
high-metallicity regions of the IGM and that many of these absorbers
are not dynamically associated with the galaxy closest in projection
on the sky.  They find that many of the absorbers could have
originated from a different galaxy at an earlier epoch and show (see
their Figure~15) that weak {\OVI} absorbers could arise between $1
\leq R/R_{\rm vir} \leq 10$, where $R$ is the galactocentric 
distance.  This range corresponds to $100~{\rm kpc} \leq R \leq 1~{\rm
  Mpc}$, depending upon the galaxy mass.

If we assume that clouds are not dynamically associated with their
identified host galaxy and that the absolute velocity offset between a
cloud and the ``host'' galaxy, $\left| \Delta v \, \right|$, is due to
Hubble flow (with zero peculiar velocity), we can estimate $R$.  The
observer line of sight Hubble flow velocity\footnote{The radial Hubble
  flow velocity of a source at $z_s$ for an observer at $z_o$ is
  $v_{\hbox{\tiny HF}} = cE(z_o)D_c(z_o,z_s)/(1+z_o)$, where
  $D_c(z_o,z_s)$ is their radial comoving separation.} for a cloud
with impact parameter $D$ at galactocentric distance $R$ from a galaxy
at $z_{\rm gal}$ is
\begin{equation}
v_{\hbox{\tiny HF}}({\rm los}) = H_0 E(z_{\rm gal}) R \sqrt{1-(D/R)^{2}} \, ,
\label{eq:hflow}
\end{equation}
where $E(z) = \sqrt{\Omega_m (1+z) + \Omega_\Lambda}$.  In
Figure~\ref{fig:hflow}(a), we plot $R$ as a function of
$v_{\hbox{\tiny HF}}({\rm los})$ for $z_{\rm gal}=0.15$ and $z_{\rm
  gal} = 0.65$, which bracket the redshifts of our sample.  When $D/R
\ll 1$, the curves are independent of $D$.

In Figure~\ref{fig:hflow}(b), we plot $R$ as a function of $\left|
v_{\hbox{\tiny HF}}({\rm los})\right|/v_{\rm esc}$ for the individual
clouds in our sample by assuming that the observed velocity offset of
the cloud is pure line of sight Hubble flow, i.e., $\left| \Delta v \,
\right| = \left| v_{\hbox{\tiny HF}}({\rm los})\right|$. Points are
colored by their $D/R_{\rm vir}$ location using the color scheme
employed for Figures~\ref{fig:Vesc} and \ref{fig:MassEscape}.  Filled
points are clouds for which {\OVI} absorption is detected with the
{\HI} absorption, and open points are {\HI} only clouds.

If the {\HI} and {\OVI} absorbing clouds are interpreted in the
context of the predictions of \citet{Oppenheimer2009} [see their
  Figure~15], we would expect the clouds reside within $R\leq 1$~Mpc
of the galaxies.  Figure~\ref{fig:hflow}(b) shows that, if the
velocity offsets are due to Hubble flow, then 31 of 45 ($\sim\! 70$\%)
absorbing clouds would be predicted to reside between 1~Mpc and
11~Mpc, and 13 of 24 ($\sim\! 50$\%) clouds with {\OVI} absorption
would be predicted to reside between 1~Mpc and 6~Mpc.  Our {\OVI}
absorbing clouds, which have $\log N({\OVI}) \simeq 14$, would not be
analogues of the ``dynamically unassociated'' {\OVI} absorbers of
\citet{Oppenheimer2009} if they reside at $R>1$~Mpc from their nearest
projected galaxies.

The same conclusions can be inferred from Figure~\ref{fig:hflow}(c),
in which we plot $R/R_{\rm vir}$ as a function of $\left|
v_{\hbox{\tiny HF}}({\rm los})\right| /v_{\rm esc}$.  Using AMR
cosmological simulations, \citet{Cuesta2008} showed that the influence
of the halo gravitational potential on dark matter particles extends
no farther than $R \simeq 4R_{\rm vir}$ for halo masses ranging from
$10 \leq \log M_{\rm\,h}/M_{\odot} \leq 14$; Hubble flow dominates for
$R/R_{\rm vir} > 4$ regardless of halo mass.  As seen in
Figure~\ref{fig:hflow}(c), if the velocity offsets are due to Hubble
flow, then $\sim\! 70$\% of the clouds and $\sim\! 60$\% of the clouds
with {\OVI} absorption reside at $R/R_{\rm vir} > 4$.  The 16 clouds
with $\left| v_{\hbox{\tiny HF}}({\rm los})\right| /v_{\rm esc} > 1$
would reside between $15 \leq R/R_{\rm vir} \leq 130$, and the 5 of
these with {\OVI} absorption would reside between $15 \leq R/R_{\rm
  vir} \leq 60$.

Based upon the above exercise, we must either adopt the interpretation
that (1) the velocity offsets of more than half of the absorbing
clouds in our sample are explained by Hubble flow and that they are
IGM absorbers at Mpc distances well beyond $R = 15R_{\rm vir}$ from
their identified galaxies, or (2) the clouds are in fact associated
with their host galaxies and that the velocity offsets are peculiar
velocities due to physical and dynamical processes within $R \simeq 4
R_{\rm vir}$.  As we discuss below, our exercise leaves very little
room for ambiguity between these very different scenarios.

The surveys of \citet{Tripp2001, Tripp2006}, and \citet{Tumlinson2005,
  Tumlinson2011} have shown that the nearest projected neighboring
galaxies are within 200~kpc of {\OVI} absorbers.  \citet{Stocke2006}
finds that, for $\log N({\OVI}) \geq 13.2$, the median distance of
{\OVI} absorbers from the nearest projected galaxy is 350-500 kpc for
$L^{\ast}$ galaxies and 200-270 kpc for $0.1~L^{\ast}$ galaxies.  In
addition, we note that if {\OVI} absorbers with column densities in
the regime detected in our survey are in the IGM at Mpc distance from
galaxies, then the covering fraction of {\OVI} absorbers should have
no dependence on galaxy property.  However, precisely the opposite is
observed in that \citet{Tumlinson2011} reports a reduced frequency of
{\OVI} absorbers in the vicinity of galaxies with lower specific star
formation rates, at least for $\log N({\OVI}) \geq 14.2$ within
$D=150$~kpc.  Finally, we argue that, if the transition from the CGM
to the IGM occurs at an overdensity of $\log \rho_{\hbox{\tiny
    H}}/\bar{\rho}_{\hbox{\tiny H}} \simeq 0.5$, as indicated by the
observations of \citet{Steidel2010}, \citet{Prochaska2011}, and
\citet{Rudie2012} and cosmological simulations such as those of
\citet{Dave1999}, then the IGM at $z< 0.5$ would correspond with $\log
N({\HI}) \la 13$.  The individual clouds we are studying have $\log
N({\HI}) \sim 14$ corresponding to $\log \rho_{\hbox{\tiny
    H}}/\bar{\rho}_{\hbox{\tiny H}} \simeq 1.3$, suggesting that they
reside in the regime of $R/R_{\rm vir} \la 4$ \citep{Klypin2011}.  We
conclude that it is unlikely that the absorbers in our sample reside
at Mpc distances in the IGM or that their velocity offsets are due to
Hubble flow.

Assuming the absorbing clouds are under the influence of the halos of
their host galaxies, we still cannot directly distinguish whether the
clouds are outflowing or inflowing from the data themselves.  However,
simple gravitational energy conserving physical arguments can be
invoked to show that infalling material is not expected to have
velocity offsets with respect to the galaxy that exceed the halo
escape velocity.  First, material does not fall into halos from
infinity, but from the ``Eulerian sphere'', a region with a $\sim\!2$
Mpc comoving radius that is set by the infall times being shorter than
the Hubble time.  Dark matter only $\Lambda$CDM cosmological
simulations support such expectations
\citep[e.g.,][]{Cuesta2008}. Though the infall velocities of a
non-negligible fraction of the infalling dark matter particles exceed
the circular velocity at the virial radius, virtually none exceed the
escape velocity.  Second, gas experiences hydrodynamic forces that act
to decelerate infalling gas.

\begin{figure}[bht]
\epsscale{1.2}
\plotone{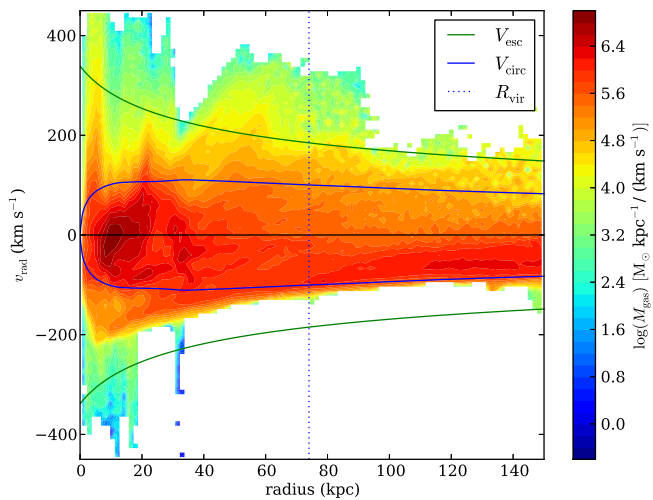}
\caption{Radial velocity versus galactocentric radius of gas mass in units
  of solar masses per unit kpc per unit velocity (see the color bar)
  showing the outflow (positive velocities) and inflow (negative
  velocities) into a simulated galaxy with $M_{\rm\,h} \simeq 2\times
  10^{11}$~M$_{\odot}$.  The vertical dotted line is the virial
  radius.  The solid green curves represent the escape velocity as
  computed from Equation~\ref{eqn:vesc} and the blue curves represent
  the circular velocity.  Note that the velocity of infalling gas does
  not exceed the escape velocity.}
\label{fig:inorout}
\end{figure}

The kinematic behavior of gas can be examined in hydrodynamic + N-body
$\Lambda$CDM cosmological simulations that compare various stellar
feedback recipes \citep{Trujillo2014}.  We examined the gas kinematics
in a simulated galaxy from the work of \citet{Trujillo2013}.  We use
the model spRP\_40, which has virial mass $\log M_{\rm\,h}/M_{\odot} =
11.3$, near the median mass of our sample.  The stellar feedback
recipe includes thermal energy from supernovae, shocked stellar winds,
and radiation pressure.  Full details can be found in
\citet{Trujillo2013}.

For this simulated galaxy, we computed the gas mass in solar masses
per unit kpc per unit velocity [{\kms}] as a function of radial
velocity and galactocentric radius and present the results in
Figure~\ref{fig:inorout}.  The dotted vertical line provides the
location of the virial radius and the solid curves provide the
circular velocity and the escape velocity as a function of
galactocentric distance.  Outflowing material (positive velocities) is
seen to have radial velocities exceeding the escape velocity, whereas
inflowing material (negative velocities) is always infalling with
radial velocities roughly a factor of $\simeq\! 1.4$ smaller than the
escape velocity at all galactocentric radii.  Outside the virial
radius, the infall kinematics reflect the circular velocity, whereas
inside the virial radius, the infall velocity increases relative to
the circular velocity toward smaller galactocentric distances.
Qualitatively identical results are found for all of the various
stellar feedback models developed by \citet{Trujillo2013}, indicating
that the infall kinematic behavior is independent of stellar feedback
model \citep[see][]{Trujillo2014}.

If the absorbers in our sample are associated with the identified host
galaxies, we can thus infer that clouds with velocities greater than
the escape velocity are entrained in outflowing material.  If so, our
data, as presented in Figure~\ref{fig:Vesc}, suggest that clouds
constrained to reside outside the virial radius have a higher escape
fraction than those that are not constrained to reside outside the
virial radius.

\subsection{Implications of Differential Kinematics}
\label{sec:kindiscussion}

Having argued that the absorbing clouds in our sample are best
interpreted as residing within $R \simeq 4 R_{\rm vir}$ of their
galaxies and that those clouds with $\left| v \, \right| > v_{\rm
  esc}$ are outflowing from the galaxy, we now explore the
implications for galaxy evolution in light of our finding of
differential kinematics (see Section~\ref{sec:vesc}).  

By differential kinematics, we are referring to the result in which we
find that the lower mass subsample has a smaller fraction of bound
clouds than the higher mass subsample and, for all masses, the bound
fraction decreases as $D/R_{\rm vir}$ increases.  Summarizing,
dividing the sample into lower mass and higher mass halos, we find
that for $D/R_{\rm vir} < 1$, lower mass halos have an escape fraction
of $\sim\! 65$\%, whereas higher mass halos have an escape fraction of
$\sim\! 5$\%.

One highly simplified yet possible explanation for the observed trends
in the data is that, the higher the launch velocity of a wind cloud,
the further from the galaxy it can potentially travel (assuming the
cloud destruction timescale is longer than the dynamical time).  Since
clouds with launch velocities below $v_{\rm esc}$ cannot achieve a
distance greater than their turn-around radius, as we probe further
out from the galaxy, we would naturally find that the fraction of
higher velocity clouds increases.

For our higher mass subsample, $11.5 < \log (M_{\rm\,h}/M_{\odot})
\leq 12.2$ , we note that the bound fractions for both {\HI} and
{\OVI} absorbing clouds are consistent with the findings of COS-Halos
\citep{Tumlinson2011, Tumlinson2013}.  However for our lower mass
subsample, $10.8 \leq \log (M_{\rm\,h}/M_{\odot}) < 11.5$, we find
higher escape fractions, and this holds for all $D/R_{\rm vir}$ bins.
{\it This implies that galaxies with halo masses {of\/}
$\log(M_{\rm\,h}/M_{\odot}) < 11.5$ expel a larger portion of their
winds to the IGM than do higher mass galaxies.  It also implies that
wind recycling would characteristically be more common in higher mass
galaxies than in lower mass galaxies.}

Thus, our result of differential kinematics has implications for the
recycling of wind material as a function of halo mass, and in fact,
is consistent with the ``differential wind recycling'' scenario
proposed by \citet{Oppenheimer2010}.  Wind recycling serves as a third
mechanism, in addition to ``cold'' and ``hot'' mode accretion
\citep[e.g.,][]{Keres2005, Dekel2006}, for gas accretion into the ISM
for fueling star formation.  In the differential wind recycling
scenario, the wind recycling time decreases with increasing halo mass,
flattening toward the highest masses.  Towards lower halo mass,
the recycling time exceeds the Hubble time, so that lower mass
galaxies would not experience wind recycling but have their star
formation fueled primarily through cold mode accretion.  That is, the
higher the halo mass, the shorter the recycling time of the wind
material, where recycling time is the sum of the time in the wind, the
time infalling back into the ISM, and the time before the gas is
incorporated into stars or expelled (again) from the ISM.  An
important factor in this behavior of winds is the higher efficiency of
hydrodynamic deceleration of the wind in higher mass galaxies due to
their larger reservoir of CGM gas and higher overdensity environments
\citep{Oppenheimer2008}.

If differential kinematics is an observational signature of
differential wind recycling in real galaxies, it would provide direct
evidence supporting the findings of \citet{Oppenheimer2010} that the
shape of the low mass end of the galaxy stellar mass function is
governed primarily by a decrease in wind recycling as halo mass
decreases.

Differential kinematics would then also provide insights into the
mass-metallicity relationship of galaxies \citep{Tremonti2004}.  If
the ISM of lower mass halos is supplied primarily via cold-mode
accretion, then the ISM of lower mass galaxies will have lower
metallicity.  On the other hand, differential kinematics implies that
the higher mass galaxies re-accrete their enriched gas, such that the
ISM enrichment of higher mass galaxies will be higher.  The idea that
outflows must be more efficient at removing metals from low-mass
galaxies is required in order for models to reproduce the observed
mass-metallicity relation has also been argued (with far more
sophistication) by several others \citep[e.g.,][]{Dalcanton2007,
  Finlator2008, Peeples2011}.

Additional observations to better characterize differential kinematics
in the CGM would be highly useful for constraining such models and
increasing our understanding of the stellar mass function and
mass-metallicity relationship.

\subsection{Comparing to Wind Models}

Here, we undertake an exercise to estimate the degree to which the
realization of the data from our sample may be consistent with simple
wind models.  We investigate three constant-velocity wind models
using the Monte Carlo technique.  We employ the two-dimensional
distribution of data presented in Figure~\ref{fig:MassEscape}. i.e.,
$\left| \Delta v \right| /v_{\rm esc}$ vs.\ $M_{\rm\, h}$, to
constrain whether the models are statistically inconsistent or are not
inconsistent with the realization of our data.  Our aim is to
determine the degree to which the paucity of data points with $\left|
v({\rm los}) \right|/v_{\rm esc} > 1$ at higher halo mass and with
$\left| v({\rm los}) \right|/v_{\rm esc} < 1$ at lower halo mass may
be a chance realization for constant velocity wind models (those not
based upon differential kinematics).

The three different wind models we investigate are: (1) a constant
outflow wind velocity, $v_{\rm w}$, independent of galaxy halo mass,
(2) a random wind velocity ranging from 0~{\kms} to a maximum
velocity, $v_{\rm w}$, also independent of galaxy halo mass, and (3)
the $vzw$ wind model of \citet{Oppenheimer2010}, in which the wind
velocity scales with the stellar velocity dispersion
$\sigma_{\ast}(r)$, where $r$ is the radius at which the winds are
launched.  The $vzw$ wind model is therefore halo mass dependent.

The model is one dimensional in which the wind velocities are
plane-parallel and randomly oriented at some angle, $\theta$, with
respect to the observer's line of sight.  There is significant
uncertainty concerning the orientation of galactic winds with respect
to galaxy inclination and position angle (i.e. orientation on the
sky).  Therefore, we adopt this simple model using an unweighted
distribution of random angles in hopes of capturing the stochastic
effects of varying galactic outflow conditions without introducing
extra free parameters and/or possible model biases.

For the first two models, we varied $v_{\rm w}$ over the range
100 to 1500~{\kms}.  For the $vzw$ model (also known as `` momentum driven
winds''), the wind velocity is given by
\begin{equation}
v_{\rm w} = 3\,\sigma_{\ast}(r) \, \sqrt{f_{\hbox{\tiny L}} - 1} \, ,
\label{eq:vzw}
\end{equation}
where $f_{\hbox{\tiny L}}$ is the luminosity factor.  Following
\citet{Oppenheimer2010}, we adopt $f_{\hbox{\tiny L}} = 2$.  The
stellar velocity dispersion at the radius where the winds are launched
is given by
\begin{equation}
\sigma_{\ast}(r) = \sqrt{-\frac{1}{2}\Phi(r)} \, ,  
\end{equation}
where $\Phi(r)$ is the gravitational potential evaluated at the 
wind launch radius.  We assume an NFW profile for which
\begin{equation}
\Phi(r) = -4 \pi G \rho_{0} r_s^2 
\,  \frac{\ln (1 + r/r_s)}{r/r_s} \, ,
\end{equation}
where $r_s$ is the scale radius, and $\rho_{0}$ is given by
\begin{equation}
\rho_0 = \frac{M(R_{\rm vir})} {4\pi r_s^3 [\ln (1+c) - c/(1+c)]} \, ,
\end{equation}
where $c$ is the concentration parameter.  Note that $M(R_{\rm vir})$
corresponds to our measurement $M_{\rm\, h}$.  The
concentration parameter is both halo mass and redshift
dependent; for this exercise, we adopt the median redshift of the
sample, $z=0.21$.

For each wind in the Monte Carlo simulation, we first generate an
associated galaxy halo mass, $M_h$, in the range of the sample
galaxies, $10.8 < \log (M_h/M_{\odot}) < 12.2$, from which we compute
the virial radius $R_{\rm vir}$.  We then generate a wind orientation
in the range $0^{\circ} < \theta < 90^{\circ}$ and an impact parameter
in the range $57 < D < 292$~kpc (the range of the sample).  We then
compute the escape velocity at the galactocentric distance equal to
the impact parameter $D$, reproducing the $v_{\rm esc}$ employed for
Figure~\ref{fig:MassEscape}.  For the constant velocity wind model, we
assign a value to $v_{\rm w}$.  For the random velocity model, we
assign a maximum value of $v_{\rm w}$ and then multiply by a random
$U(0,1)$ deviate.  In the case of the momentum-driven $vzw$ wind
model, we specify the launch radius of the wind and compute $v_{\rm
  w}$ from Equation~\ref{eq:vzw}.  Finally, we determine the line of
sight ``observed'' velocity $\left| v({\rm los}) \right| = v_{\rm w}
\cdot \cos(\theta)$, from which we compute the ratio $\left| v({\rm
    los}) \right| /v_{\rm esc}$.

For a given wind model, we generate 100,000 realizations (galaxy/wind
pairs).  From these pairs, we randomly draw 41 galaxy/wind pairs but
enforce that the 41 pairs match the number of data points on
Figure~\ref{fig:MassEscape} with $D/R_{\rm vir} \leq 1$, $1 < D/R_{\rm
  vir} \leq 2$, $2 < D/R_{\rm vir} \leq 3$ in each of four equally
spaced halo mass bins over the range $10.8 \leq \log M_{\rm\,
  h}/M_{\odot} \leq 12.2$.  Thus, the two-dimensional distribution of
halo mass and $D/R_{\rm vir}$ of the 41 galaxy/wind pairs emulates
that of the observed data on Figure~\ref{fig:MassEscape}.  On the
$\left| v({\rm los}) \right| /v_{\rm esc}$--$M_{\rm\, h}$ plane, we
then compute the two-dimensional KS statistic between the galaxy/wind
pairs and the data points in order to quantify the degree to which the
distribution of wind model points is inconsistent with the
distribution of observed points.  We adopt the criterion that the
model points are inconsistent with the data when $P({\rm KS}) \leq
0.0027$, corresponding to a 99.97\% ($3~\sigma$) or higher confidence
level.

We repeat the entire process for 100,000 trials, each time calculating
the two-dimensional KS probability comparing the model data to the
observed data.  Finally, we compute the fraction of model trials for
which $P({\rm KS}) \leq 0.0027$ (i.e., the fraction out of 100,000 for
which the model data can be ruled inconsistent with the
observed data at the $3~\sigma$ level).  As this fraction,
$f(P_{\hbox{\tiny KS}} \!<\! 0.0027)$, approaches unity, the wind
model is less consistent with the data.

\begin{figure}[bth]
\epsscale{1.2}
\plotone{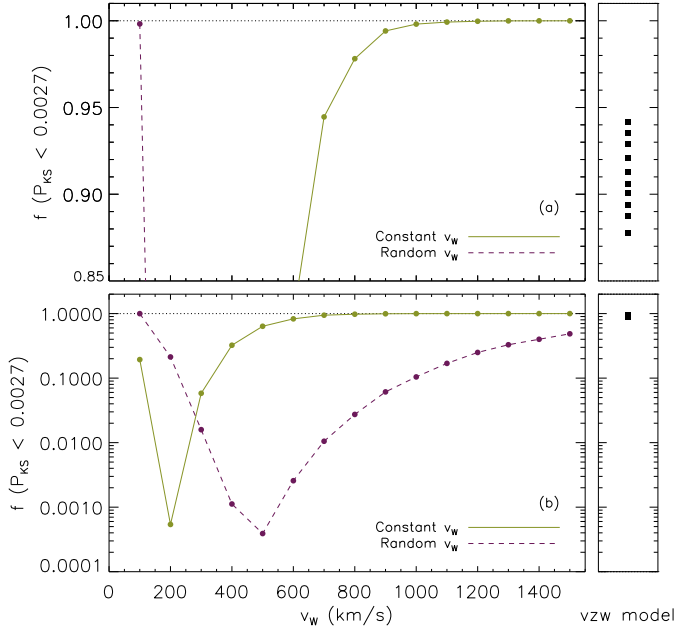}
\caption{The fraction of realizations, $f(P_{\hbox{\tiny KS}} \!<\!
  0.0027)$, that are inconsistent with the data at the $3~\sigma$
  level versus the wind velocity, $v_{\rm w}$, for three different wind
  models.  Panel (a) shows a linearly scaled zoom-in of the upper
  region of panel (b).  For the constant wind velocity model, $v_{\rm
    w}$ is the wind velocity.  For the random wind velocity model,
  $v_{\rm w}$ is the maximum wind velocity, which can range from
  $0$~{\kms} to $v_{\rm w}$.  The $vzw$ wind model (right panels) is
  computed for wind launch radii over the range $r=1$--10~kpc in steps
  of 1~kpc.  The upper point corresponds to $r=1$~kpc and the lower
  point corresponds to $r=10$~kpc.  The random wind models with wind
  velocities peaking around $\sim500$~{\kms} are most frequently
  consistent with the data.}
\label{fig:windmodels}
\end{figure}

In Figures~\ref{fig:windmodels}(a) and \ref{fig:windmodels}(b), we
plot $f(P_{\hbox{\tiny KS}} \!<\!  0.0027)$ as a function of $v_{\rm
  w}$ for the constant velocity and random velocity wind models.  For
the random wind model, $v_{\rm w}$ represents the maximum value of the
constant velocity wind.  Panel (a) is a linearly scaled zoom-in of the
upper portion of panel (b).  Where the curves have $f(P_{\hbox{\tiny
    KS}} \!<\!  0.0027) \ge 0.9$, the distribution of the model
data is ruled out at the $3~\sigma$ level for 90\% or more of the
realizations.  The right-hand panels show $f(P_{\hbox{\tiny KS}} \!<\!
0.0027)$ for the $vzw$ wind model for ten different launch radii
ranging from $r=1$ to 10~kpc in steps of 1~kpc.  For this model, the
value of $f(P_{\hbox{\tiny KS}} \!<\!  0.0027)$ increases as the
launch radius decreases (the highest $f(P_{\hbox{\tiny KS}} \!<\!
0.0027)$ corresponds to $r=1$ kpc).

For the constant velocity wind model, $f(P_{\hbox{\tiny KS}} \!<\!
0.0027) \ge 0.9$ occurs for $v_{\rm w} \ge 650$~{\kms}.  That the
range of velocities below this value is less frequently inconsistent
with the data is not outside of expectations, since 90\% of the {\HI}
absorbing cloud velocities lie within $\Delta v = \pm 500$~{\kms}.
Note that it is very rare for the realizations to be inconsistent with
the data for $v_{\rm w} \simeq 200$~{\kms}, the value within which
70\% of all {\HI} absorbing cloud velocity offsets lie with respect to
the galaxy.

For the random velocity wind model, $f(P_{\hbox{\tiny KS}} \!<\!
0.0027) \ge 0.9$ occurs for $v_{\rm w} \simeq 100$~{\kms}.  Recall
that in this model, the wind velocity of any given galaxy/wind pair
falls in the range 0~{\kms} to the maximum velocity, $v_{\rm w}$.  We
find $f(P_{\hbox{\tiny KS}} \!<\!  0.0027) < 0.9$ occurs for all
wind velocities $v_{\rm w} > 100$~{\kms}, indicating that the random
velocity wind model cannot be ruled as being inconsistent with the
data for more than 90\% of the realizations over the range $100 <
v_{\rm w} \leq 1500$~{\kms}.  

For the $vzw$ wind model, $f(P_{\hbox{\tiny KS}} \!<\!  0.0027) \ge
0.90$ occurs for launch radii $r \leq 5$~{kpc}.  Formally, the $vzw$
wind model cannot be ruled as inconsistent with the data; however, for
reasonably physical launch radii, 90\% of the realizations are
inconsistent with the data to the $3~\sigma$ level.  Generally, the
$vzw$ wind model does not exhibit a compelling signature for being
consistent with the data.

That the random velocity wind model can be consistent with the data
suggests that, even in a small sample (14 galaxies with 41 absorbing
clouds), highly variable wind velocities from galaxy to galaxy can
give rise to the differential kinematics we inferred from the data.
This remains consistent with our previous statement that one possible
explanation for differential kinematics is that (in real galaxies) the
higher the launch velocity of the wind, the further from the galaxy
the absorbing clouds potentially travel, so that as we probe further
out from a galaxy we observe a higher fraction of higher velocity
clouds.

If the cloud velocities are decelerated dynamically in higher mass
galaxies, as found in the simulations of \citet{Oppenheimer2010}, then
higher wind velocities would be more frequently observed in the outer
extended CGM of lower mass galaxies, as we have inferred for our
sample.  In the wind model of \citet{Chelouche2010}, the wind velocity
is proportional to the star formation rate in the galaxy disk.  We do
not have estimates of the star formation rates for the galaxies in our
sample.  But, we note that variations in the star formation rate from
galaxy to galaxy would manifest in their model in a manner similar to
our random velocity wind model.  We also note, generally, that lower
(stellar) mass galaxies tend to have higher specific star formation
rates than higher mass galaxies \citep[cf.,][]{Schawinski2014}, and it
seems reasonable that specific star formation rate would correlate
with wind velocity.

%=============== CONCLUSIONS ================================

\section{Conclusions}
\label{sec:conclusion}

We have presented an analysis of the spatial and geometric
distribution and kinematics of {\HI} and {\OVI} absorption surrounding
14 galaxies within a projected distance of $D = 300$~kpc of background
quasars. The galaxies are imaged using {\it HST}/WFPC2 and their
morphological and orientation parameters have been measured using
GIM2D.  The absorption is measured in {\it HST}/COS or {\it HST}/STIS
quasar spectra.  We have focused our analysis on the {\Lya} and {\Lyb}
transitions, and {\OVIdblt} doublet.  The column densities, number of
clouds, and the kinematics were measured using Voigt profile fitting.

The sample is characterized by a redshift range $0.12 \leq z \leq
0.67$, and an impact parameter range $60 \leq D \leq 290$ kpc.  The
galaxy virial masses range from $10.8 \le \log(M_{\rm\,h}/M_{\odot})
\le 12.2$, corresponding to virial radii between $70 \le R_{\rm vir}
\le 225$~kpc.  The median virial mass is $\log(M_{\rm\,h}/M_{\odot}) =
11.5$.  The range of $D/R_{\rm vir}$ spans from $0.45$ to $2.75$.  The
range of galaxy inclinations and azimuthal angles are $18^{\circ} \le
i \le 85^{\circ}$ and $6^{\circ} \le \Phi \le 87^{\circ}$,
respectively.

\subsection{Spatial and Geometric Distributions}

We first highlight some general results with regards to the spatial
and geometric distribution of {\HI} and {\OVI} absorbing gas in the
CGM.

1. Although higher $N({\HI})$ systems are found at $D<100$~kpc, there
is no statistical trend between $N({\HI})$ and $D$.  Over the range
$100 \leq D \leq 300$~kpc, the system total $N({\HI})$ is typically
$\log N({\HI}) \simeq 14$.  For all $D$, the mean and dispersion is
$\log \langle N({\HI}) \rangle = 14.34 \pm 0.61$.  We detect {\OVI} as
far as $D\sim 290$ kpc (our sample maximum), to a $3~\sigma$ limit of
$\log N({\OVI}) = 12.8$.  The distribution of $N({\OVI})$ is
effectively flat as a function of $D$, showing no statistically
significant trend between $N({\OVI})$ and projected distance from the
host galaxy.  The mean and dispersion is $\log \langle N({\OVI})
\rangle = 14.03\pm0.44$.

2. There is a higher average value and a broader spread in $N({\HI})$
for $D/R_{\rm vir} < 1$ as compared to $D/R_{\rm vir} > 1$, with $
\log \langle N({\HI}) \rangle = 15.9 \pm 1.6$ inside and $\log \langle
N({\HI}) \rangle = 14.1 \pm 0.5$ outside the projected virial radius,
respectively.  Due to the flat spatial distribution of $N({\OVI})$
with $D$ and the higher average value and dispersion of $N({\HI})$ at
$D/R_{\rm vir} < 1$, the dispersion in $N({\OVI})/N({\HI})$ is a
factor of $\simeq 8$ greater inside the projected virial radius.

3. The is no discernible trend between the system total $N({\HI})$,
$N({\OVI})$, or $N({\OVI})/N({\HI})$ and galaxy inclination.
Statistically, there is no correlation between these quantities and
azimuthal angle.  However, in our small sample, $N({\HI})$ is largest
when probed nearest to the project axes of the galaxy and decreases as
the azimuthal angle increases away from the projected axes.

4. Within $D=300$~kpc, there is a higher incidence of {\OVI}
absorption in higher mass halos than in lower mass halos, using
the sample median of $\log M_{\rm\,h}/M_{\odot} = 11.5$ to divide the
masses into ``higher'' and ``lower''.  We find associated {\OVI}
absorption in only $\sim\!40$\% of the {\HI} clouds in and around
lower mass halos as compared to $\sim\!85$\% around higher mass halos.
Since the system total $N({\HI})$ is fairly flat, the smaller fraction
of detected {\OVI} clouds in lower mass halos suggest conditions
favoring {\OVI} is less common out to $D = 300$ kpc of lower mass
halos than for higher mass halos, but that the physical conditions of
the gas are not dissimilar.

In summary, the highest $N({\HI})$ clouds reside within the projected
virial radius and are found at azimuthal angles closely aligned with
the galaxy projected axes.  It could be that for $D/R_{\rm vir} < 1$,
the {\HI} in our sample is reflecting the presence of a cool/warm gas
phase preferentially found along the projected galaxy axes, such is
observed for {\MgII} absorption \citep{Bordoloi2011, Kacprzak2012-PA,
  Bouche2012}.  The data we have in hand cannot definitively address
the presence of a cool/warm phase.  Overall, it appears that there is
a transition in the behavior of {\HI} absorption in the regime of
$D/R_{\rm vir} \sim 1$, in which higher system total $N({\HI})$ is
found inside the projected virial radius and $\log N({\HI}) \simeq 14$
outside the projected virial radius at least as far as $D/R_{\rm vir}
\simeq 3$.  For {\OVI} absorption, the distribution of the system
total $N({\OVI})$ is flat for all $D/R_{\rm vir }$ out to at least
$D/R_{\rm vir} \sim 2.8$.  {\OVI} absorbers are more common in the CGM
of higher mass halos out to $D\simeq 300$ kpc.  Altogether,
$N({\OVI})$ shows no preferred geometric dependencies, suggesting that
regions of hot CGM gas are quite globally distributed.

\subsection{Differential Kinematics}

The main result of this paper is differential behavior in the fraction
of bound clouds (individual VP components) as a function of both
virial mass, $M_{\rm\,h}$, and virial radius, $R_{\rm vir}$.  We
called this behavior ``differential kinematics''.  These findings are
shown in Figure~\ref{fig:MassEscape} and Table~\ref{tab:bound}.
Figure~\ref{fig:MassEscape} shows the absolute relative velocity of
the Voigt profile ``cloud'' velocities with respect to the galaxy
normalized to the escape velocity, $\left| \Delta v /v_{\rm esc}
\right|$, as a function of virial and stellar mass, $M_{\rm\,h}$ and
$M_{\ast}$.  Table~\ref{tab:bound} lists the fraction of clouds that
can be inferred to be bound to the host halo as a function of
$M_{\rm\,h}$ and $D/R_{\rm vir}$. The interpretation relies heavily
upon the inference (presented in Section~\ref{sec:inorout}) that
clouds with $\left| \Delta v /v_{\rm esc} \right| > 1$ are most
probably outflowing through the CGM, and are not IGM Hubble flow or
infalling CGM or IGM clouds.  We highlight the results from our
kinematic analysis.

1. Most {\HI} and {\OVI} absorbing clouds are clustered within $\sim\!
500$~{\kms}.  In $\sim\!50$\% of the systems with detected {\OVI}
absorption, we observe a velocity offset between the bulk of the {\HI}
and the bulk of the {\OVI}, as defined by the highest column density
Voigt profile components in a system. In three of these cases, the
velocity offset is $\sim\!100$~{\kms}.  The data support the idea that
the $\log N({\HI})\simeq 14$ regime of the CGM represents various gas
conditions as inferred from {\HI} and {\OVI} absorption, even though
the system {\it total\/} $N({\HI})$, $N({\OVI})$, and
$N({\OVI})/N({\HI})$ show little variation from system to system.

2. When the full range of $M_{\rm\,h}$ and $D/R_{\rm vir}$ of the
sample are examined, $\sim\!40$\% of the {\HI} absorbing clouds can be
inferred to be escaping their host halo.  Segregating the sample into
finite ranges of $D/R_{\rm vir}$, the fraction of bound clouds
decreases as $D/R_{\rm vir}$ increases such that the escaping fraction
is $\sim\!15$\% for $D/R_{\rm vir} < 1$, $\sim\!45$\% for $1
\leq D/R_{\rm vir} < 2$, and $\sim\!90$\% for $2 \leq D/R_{\rm vir} <
3$.  That is, averaged over all $M_{\rm\,h}$, the fraction of {\HI}
absorbing clouds that could be escaping the galaxy increases with
increasing $D/R_{\rm vir}$.

3. Dividing the sample into lower mass and higher mass halos, where
the dividing virial mass is the median of the sample, $\log
M_{\rm\,h}/M_{\odot} = 11.5$, we find that the lower mass subsample
has a smaller fraction of bound clouds in each of the three
aforementioned $D/R_{\rm vir}$ ranges.  For $D/R_{\rm vir} < 1$,
lower mass halos have an escape fraction of $\sim\!65$\%, whereas
higher mass halos have an escape fraction of $\sim\!5$\%.  For $1 \leq
D/R_{\rm vir} < 2$, the escape fractions are $\sim\!55$\% and
$\sim\!35$\% for lower mass and higher mass halos, respectively.  For
$2 \leq D/R_{\rm vir} < 3$, the escape fraction for lower mass halos
is $\sim\!90$\% (higher mass halos were not probed in this range in
our sample).

4. We demonstrated that the absorbing clouds are likely to be
outflowing winds, since their kinematics are not consistent with
infall kinematics, based upon feedback simulations.  We showed that
the absorbing gas is likely to reside within $4R_{\rm vir}$ of the
galaxies, also based upon simulations and the dynamics of Hubble flow.
We explored three constant velocity wind models to explore the degree
to which the observed characteristics of differential kinematics are
inconsistent with these models.  We find that the most consistent
constant wind velocity model is that with random winds velocities in
the range $300 \leq v_{\rm w} \leq 800$~{\kms}, and suggest that
specific star formation rate, from galaxy to galaxy, coupled with
higher dynamical deceleration of the gas in higher mass halos, may be
instrumental in explaining differential kinematics.

Differential kinematics may be an observational signature supporting
the theoretical scenario of differential wind recycling proposed by
\citet{Oppenheimer2010}.  If so, differential kinematics would be an
important finding that should be verified and further characterized
with additional observations.  It is becoming well accepted that wind
recycling through the CGM is an important regulating process for
galaxy evolution and may, to a large degree, control the shape of the
stellar to halo mass function and the mass-metallicity relationship of
galaxies.

\acknowledgments

We thank Ben Oppenheimer for helpful and insightful discussions on the
details of his wind simulations.  NLM, CWC, and SM were supported
mainly through grant HST-GO-13398 and JCC and SM were partially
supported by grant HST-AR-12644, both provided by NASA through the
Space Telescope Science Institute, which is operated by the
Association of Universities for Research in Astronomy (AURA) under
NASA contract NAS 5-26555.  ST-G was supported through the Research
Enhancement Program awarded to CWC provided by NASA's New Mexico Space
Grant Consortium (NMSGC).  NLM and NMN were partially supported
through NMSGC Graduate Research Fellowships. NMN was also partially
supported through a three-year Graduate Research Enhancement Grant
(GREG) sponsored by the Office of the Vice President for Research at
New Mexico State University.

\appendix

\section{Individual Quasar Fields}
\label{sec:individualfields}

\begin{figure*}
\epsscale{1.15}
\plotone{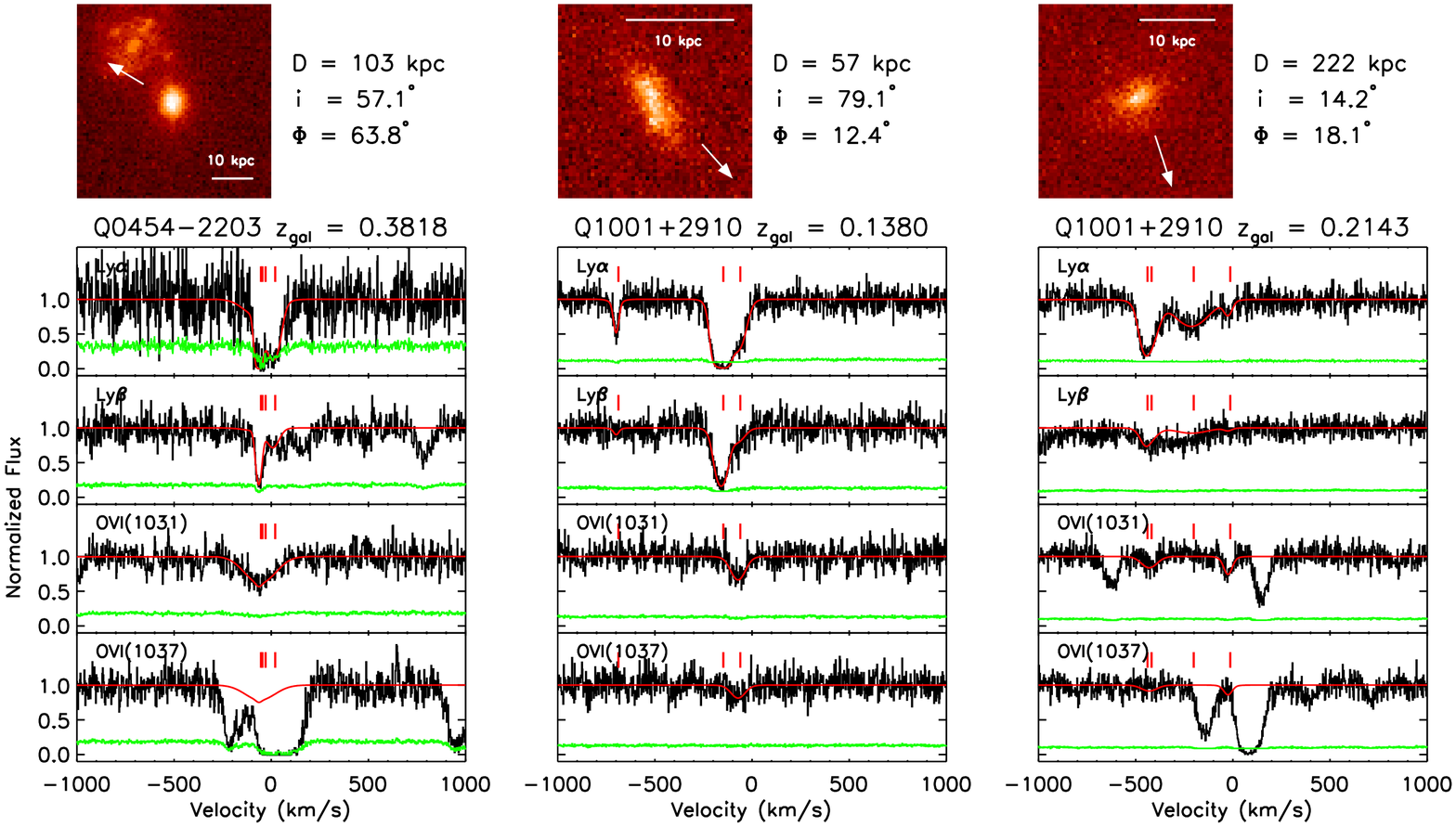}
\caption{Same as Figure~\ref{fig:profile}, but for the galaxy at
  $z_{\rm gal} = 0.3818$ in the field toward Q0454$-$2203, and the
  galaxies at $z_{\rm gal} = 0.1380$ and $z_{\rm gal} = 0.2143$ in the
  field toward Q1001$+$2910.}
\label{fig:prof2}
\end{figure*}

\begin{figure*}
\epsscale{1.15}
\plotone{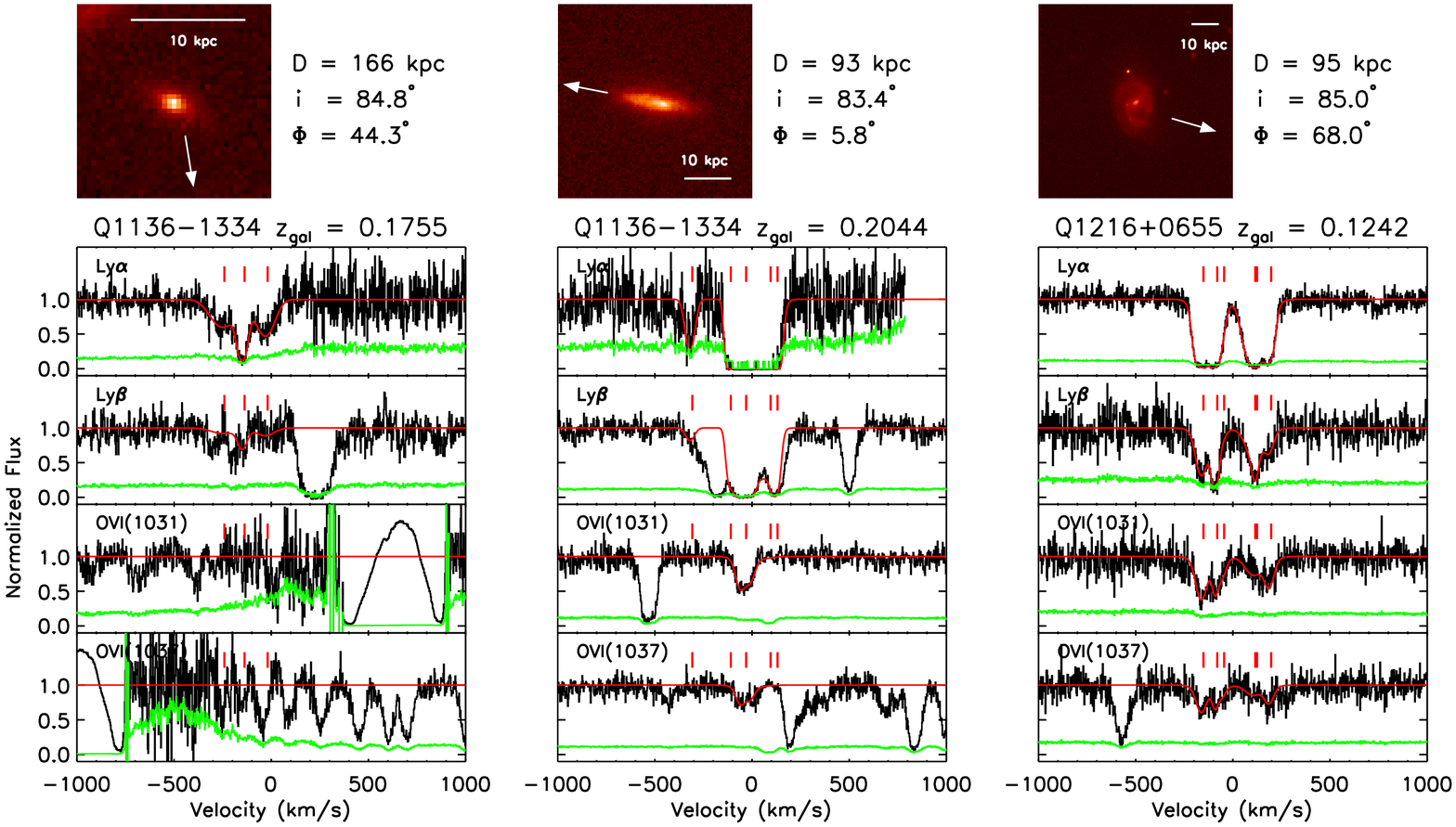}
\caption{Same as Figure~\ref{fig:profile}, but for the galaxies at
  $z_{\rm gal} = 0.1755$ and $z_{\rm gal} = 0.2044$ in the field
  toward Q1136$-$1334, and the galaxy at $z_{\rm gal} = 0.1242$ in the
  field toward Q1216$+$0655.}
\label{fig:prof3}
\end{figure*}

\begin{figure*}
\epsscale{1.15}
\plotone{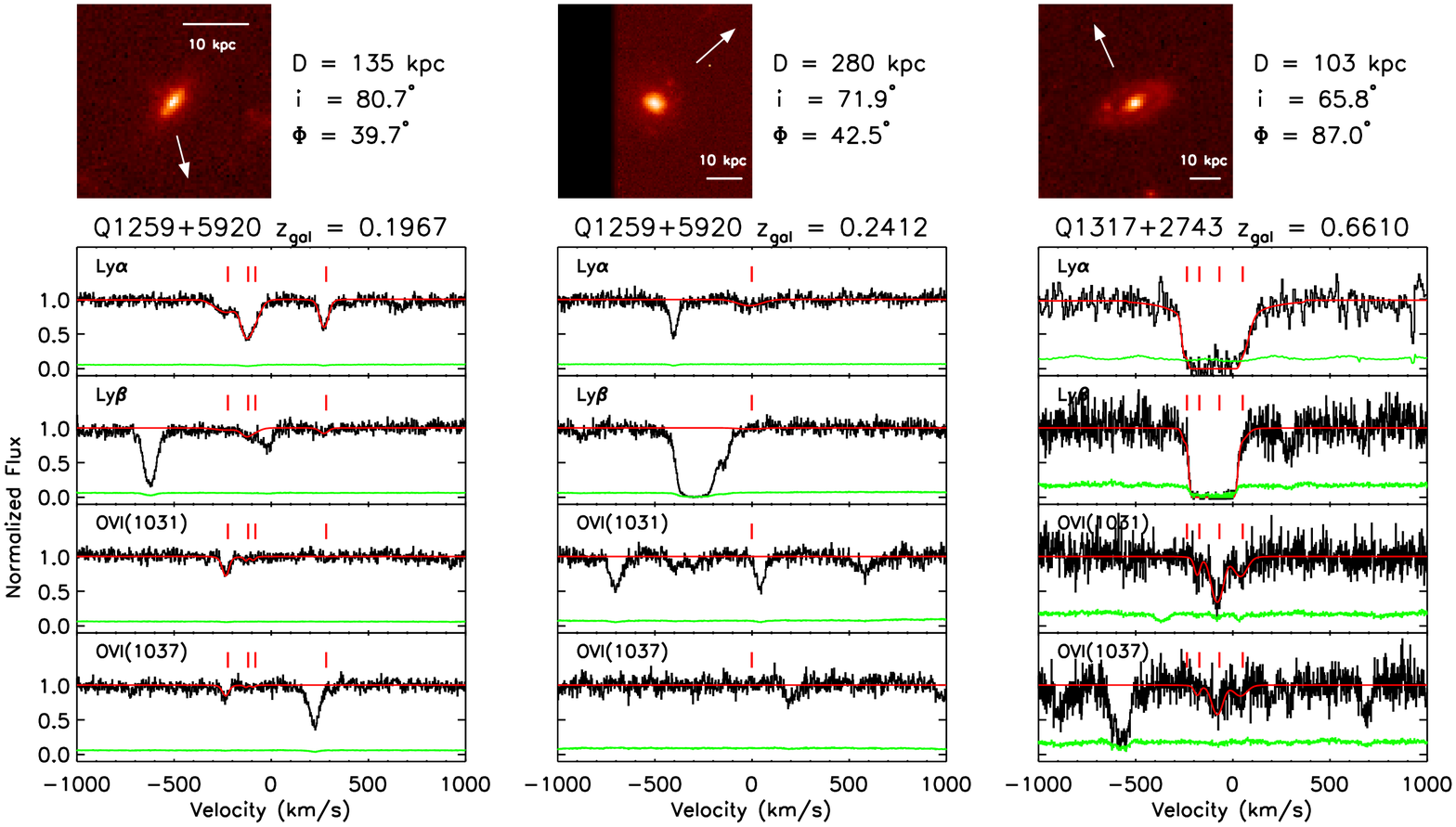}
\caption{Same as Figure~\ref{fig:profile}, but for the galaxies at
  $z_{\rm gal} = 0.1967$ and $z_{\rm gal} = 0.2412$ in the field
  toward Q1259$+$5920, and the galaxy at $z_{\rm gal} = 0.6610$ in the
  field toward Q1317$+$2743.}
\label{fig:prof4}
\end{figure*}

\begin{figure*}
\epsscale{0.8}
\plotone{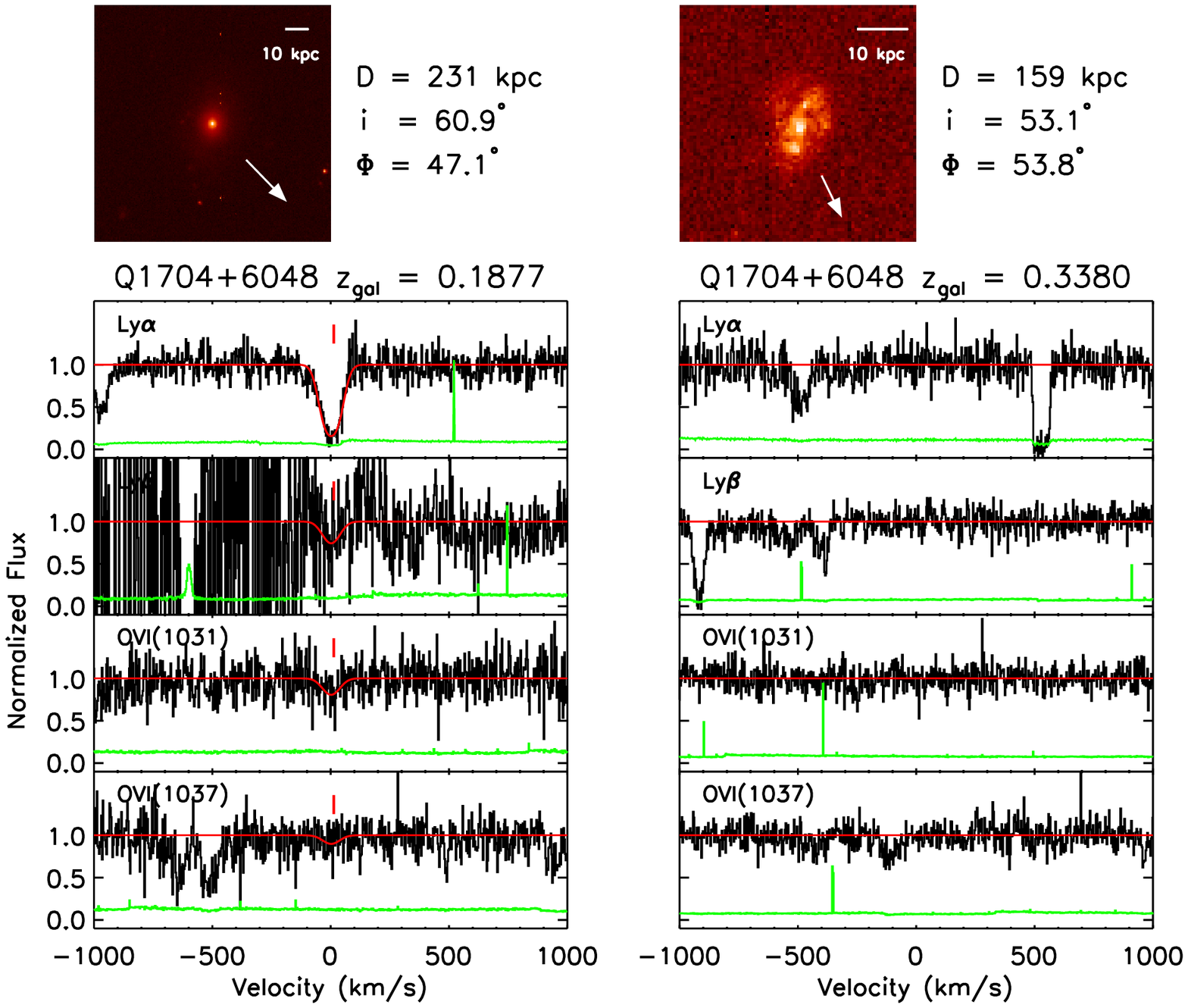}
\caption{Same as Figure~\ref{fig:profile}, but for the galaxies at
  $z_{\rm gal} = 0.1877$ and $z_{\rm gal} = 0.3380$ in the field
  toward Q1704$+$6048.}
\label{fig:prof5}
\end{figure*}

%============== INDIVIDUAL FIELDS ================================

All spectroscopic redshift data for the galaxies analyzed in this
paper come from one of five different sources. (1) Very early work was
conducted by \citet{Ellingson1994}, who employed the MARLIN/LAMA
Multiobject Spectrograph on the Canada-France-Hawaii Telescope
(CFHT). They cite a 78\% completeness level for successful
spectroscopic identification of observed galaxies with $m_r \le 21.5$,
and 49\% for the fields overall. (2) A survey by \citet{Lanzetta1995}
uses the Kast Spectrograph on the Lick Observatory 3-m telescope. This
study is 37\% complete for $m_r < 21.5$, with limiting magnitudes of
$m_r = 23.0$ and $m_r = 23.5$ for the fields toward Q1001$+$2910 and
Q1704$+$6048, respectively.  This survey adopted some of the galaxy
identifications from \citet{Ellingson1994}. (3) A survey by
\citet{LeBrun1996} was conducted with the MOS multi-slit spectrograph
on the CFHT.  They they claim a success rate of 81\% for all observed
galaxies to a limiting magnitude of $m_r = 22.5$. (4) An {\it HST}
imaging survey, using FOS quasar spectroscopy, was conducted by
\citet{Chen1998, Chen2001}, targeting the fields studied by
\citet{Lanzetta1995} and additional fields for which much of the
details are to appear in Chen et~al.~(2001, in preparation).
Estimates on the completeness and magnitude limits for quasar fields
using only these observations will be made from published data on a
field-by-field basis below. Finally, (5), \citet{Johnson2013},
performed detailed spectroscopic follow-up observations of the
galaxies in the field of Q0405$-$123. Their study employed IMACS and
LDSS3 on Magellan and DIS on the Apache Point 3.5-m telescope. They
cite a 100\% completeness level for $L > 0.1L_*$ galaxies at impact
parameters less than 100 kpc and a 75\% completeness level for $L >
0.1L_*$ galaxies at impact parameters less than 500 kpc.

Galaxy and absorber data can be found in Tables~\ref{tab:galdata} and
\ref{tab:absdata}, respectively.  Galaxy image footprints and analyzed
spectra for {\Lya}, {\Lyb}, and {\OVIdblt} profiles are show in
Figure~\ref{fig:profile} in Section~\ref{sec:CGM} and here in
Figures~\ref{fig:prof2}--\ref{fig:prof5}.

\subsection{The Field Toward Q0405$-$123}
\label{sec:Q0405}

This field was first spectroscopically surveyed by
\citet{Ellingson1994} and has had follow-up observations published in
\citet{Johnson2013}. Nearly all of the redshifts measured by
\citet{Ellingson1994} have been revised. Some galaxies associated with
absorbers in \citet{Chen2001}, which uses the Ellingson redshifts, have
changed significantly. The galaxy measured at $z_{\rm gal} = 0.3520$
was revised to $z_{\rm gal} = 0.3422$ and a galaxy located off the
WFPC2 image footprint was identified at $z_{\rm gal} = 0.3521$. The
galaxy originally measured at $z_{\rm gal} = 0.2800$ was revised to
$z_{\rm gal} = 0.7115$ and a new galaxy at $z_{\rm gal} = 0.4100$ was
added to the field. There are a large number of galaxies clustered
both spatially and in redshift space at $z_{\hbox{\tiny QSO}} = 0.57$
that are not considered in this study due to their likely physical
connection to the quasar.  In addition, there are four galaxy pairs
whose absorption cannot be disentangled (the first at $z_{\rm gal} =
0.1669$ and $z_{\rm gal} = 0.1672$, the second at $z_{\rm gal} =
0.5169$ and $z_{\rm gal} = 0.5161$, the third at $z_{\rm gal} =
0.3422$ and $z_{\rm gal} = 0.3407$, and the fourth at $z_{\rm gal} =
0.3614$ and $z_{\rm gal} = 0.3608$). We exclude these galaxies from
our analysis.  

We examine galaxy-absorber pairs at redshifts $z_{\rm gal} = 0.1534,
0.2978$, and $0.4100$.

\subsection{The Field Toward Q0454$-$2203}
\label{sec:Q0454}

The galaxy identifications come from \citet{Chen1998}, but the
spectroscopic survey of the field remains unpublished.  From
\citet{Chen1998}, we estimate that the survey limiting magnitude is
$m_r \sim 21.8$. There are two galaxies in the field within $\sim\!
300$~{\kms} of each other ($z_{\rm gal} = 0.4837$ and $z_{\rm gal} =
0.4847$) that are not included in our study (they also do not have UV
spectra covering {\Lya}, {\Lyb}, and {\OVI}).

We examine only the galaxy-absorber pair for the galaxy at $z_{\rm gal} = 0.3818$. 

\subsection{The Field Toward Q1001$+$2910}
\label{sec:Q1001}

This field was spectroscopically surveyed by
\citet{Lanzetta1995}. There are only a few bright galaxies near the
quasar, allowing straight-forward identification of galaxy-absorber
pairs. We note that the galaxy at $z_{\rm gal} = 0.2143$ does not
appear in \citet{Chen1998}, but does appear in \citet{Chen2001} with no
elaboration.  We adopt the \citet{Chen2001} data.

We examine galaxy-absorber pairs at redshifts $z_{\rm gal} = 0.1380$
and $0.2143$.

\subsection{The Field Toward Q1136$-$1334}
\label{sec:Q1136}

As with the Q0454$-$2203 field, the galaxy identifications come from
\citet{Chen1998}, but the detailed spectroscopic survey of the field
remains unpublished.  From \citet{Chen1998}, we estimate that the
survey limiting magnitude is $m_r \sim 22.3$.  There are three
galaxies clustered around $z_{\rm gal} \simeq 0.36$ which are excluded
from our study due also to a lack of UV spectral coverage of {\Lya},
{\Lyb}, and {\OVI} absorption in the COS spectrum.

We examine galaxy-absorber pairs at redshifts $z_{\rm gal} = 0.1755$
and $0.2044$.

\subsection{The Field Toward Q1216$+$0655}
\label{sec:Q1216}

The galaxy identifications come from \citet{Chen2001}, but the
detailed spectroscopic survey of the field remains unpublished.  From
\citet{Chen2001}, we estimate that the survey limiting magnitude is
$m_r \sim 21.6$.  There is only one galaxy identified that has a
redshift lower than that of the quasar with the required spectral
coverage in the COS spectrum.

We examine the only galaxy-absorber pair at redshift $z_{\rm gal} = 0.1242$. 

\subsection{The Field Toward Q1259$+$5920}
\label{sec:Q1259}

The galaxy identifications come from \citet{Chen2001}, but the detailed
spectroscopic survey of the field remains unpublished.  From
\citet{Chen2001}, we estimate that the survey limiting magnitude is
$m_r \sim 21.1$.  
 
We examine galaxy-absorber pairs at redshifts $z_{\rm gal} = 0.1967$
and $0.2412$.  These two galaxies were imaged by different programs
(see Table \ref{tab:obs}).

\subsection{The Field Toward Q1317$+$2743}
\label{sec:Q1317}

This field was spectroscopically surveyed by \citet{LeBrun1996}. They
report two pairs of galaxies in this field ($z_{\rm gal} = 0.5397$ and
$0.5398$, and $z_{\rm gal} = 0.6715$ and $0.6717$). All absorption of
interest from the $z \simeq 0.54$ galaxies falls in the gaps in the
COS spectrum.  \citet{Chen2001} also studied this field.  The most
recent work is from \citet{Churchill2012Q1317}, who studied the
$z_{\rm gal} = 0.6719$ galaxy and \citet{Kacprzak2012-1317}, who
studied the $z_{\rm gal} = 0.6610$ galaxy.  These galaxy redshifts are
revisions from the \citet{LeBrun1996} work.

Because of the complexity of the absorption associated with the
$z_{\rm gal} = 0.6719$ galaxy, we examine only the galaxy-absorber
pair at redshift $z_{\rm gal} = 0.6610$.

\subsection{The Field Toward Q1704$+$6048}
\label{sec:Q1704}

This field was spectroscopically surveyed by both \cite{Lanzetta1995}
and \cite{LeBrun1996}.

Ambiguity in this field exists for the galaxies identified at $z_{\rm
  gal} = 0.2260$ and $z_{\rm gal} = 0.2217$. Both galaxies are first
measured and identified by \cite{LeBrun1996}, but only the galaxy at
$z_{\rm gal} = 0.2260$ appears in \cite{Chen2001} because of its
proximity to the quasar line of sight ($D = 260~\mathrm{kpc}$).
\cite{Chen2001} do not identify an absorber with the $z_{\rm gal} =
0.2260$ galaxy. The other galaxy, with $D \sim 530$~kpc, has a
redshift nearly coincident with absorber at $z_{\rm abs} =
0.2216$. Due to the large transverse distance from the quasar and its
possible ambiguity with the galaxy at $z_{\rm gal} = 0.2260$, we do
not include these galaxies in our analysis.

We examine galaxy-absorber pairs at redshifts $z_{\rm gal} = 0.1877$
and $0.3380$.

% SYSTEM PLOTS ===================================================================

\section{Column Densities}
\label{sec:Lyg}

% END SYSTEMPLOTS ===========================

\begin{figure}[thb]
\epsscale{1.2}
\plotone{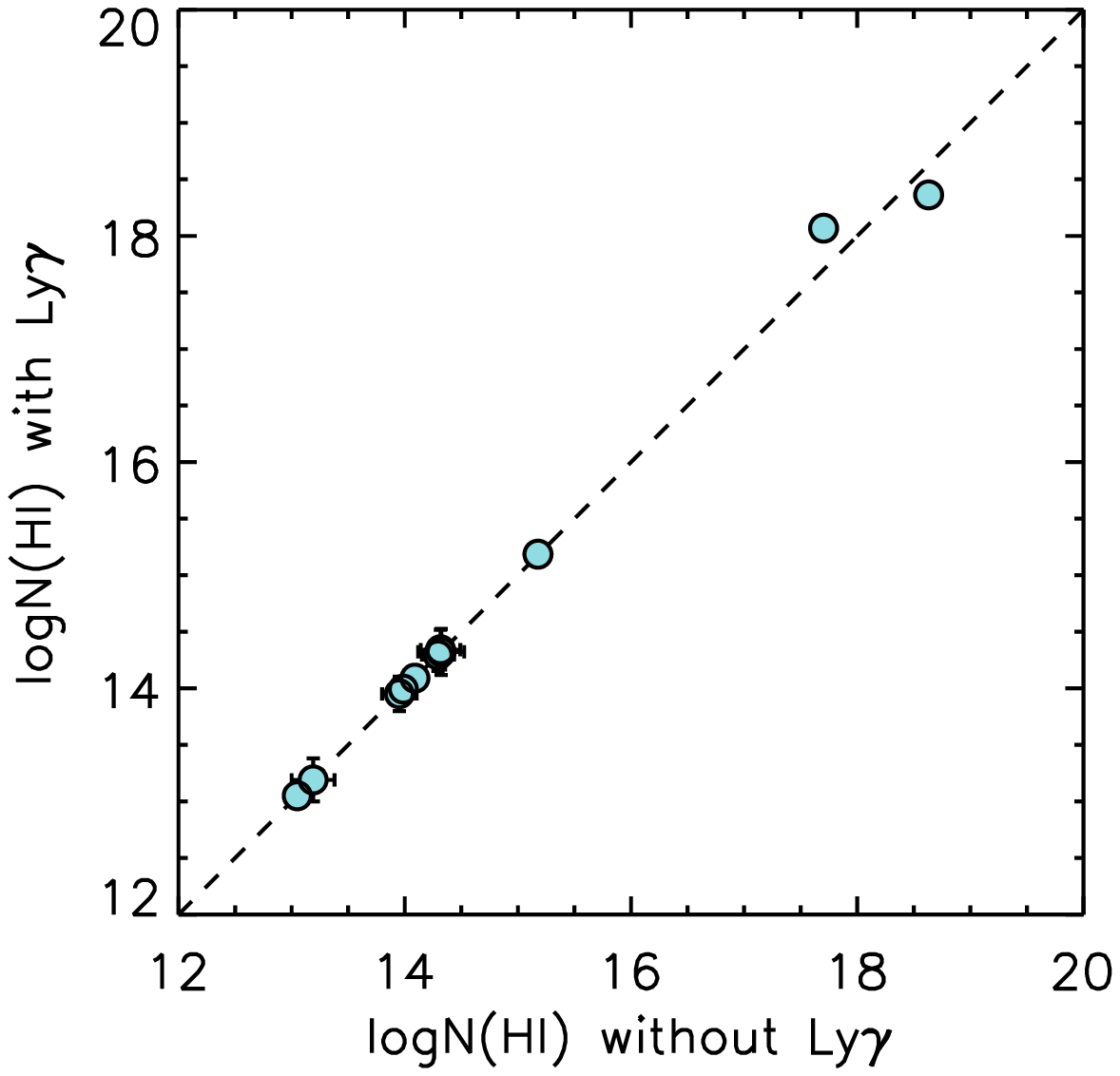}
\caption{Voigt profile column density results for $\log N({\HI})$. On the x-axis is the resultant column density measured using only {\Lya} and {\Lyb}. On the y-axis is the result using {\Lya}, {\Lyb}, and {\Lyg}. The dotted line shows a one-to-one correlation. Including the {\Lyg} with the fit has very little impact on the measured {\HI} column density. }
\label{fig_Gammacompare}
\end{figure}

For this work, we present {\HI} column densities using only the {\Lya}
and {\Lyb} transitions.  For roughly half of our sample the {\Lyg} is
also available for fitting.  As such, requiring {\Lyg} for the fits
would significantly reduce our sample size.

To ensure that the Voigt profile fits using {\Lya} and {\Lyb} only are
not systematically skewed relative to fits using {\Lya}, {\Lyb}, and
{\Lyg}, we compared the fits with and without {\Lyg} for the subsample
that has {\Lyg} coverage.

In Figure~\ref{fig_Gammacompare}, we present the {\HI} column
densities derived from {\Lya} and {\Lyb} only fits and {\Lya}, {\Lyb},
and {\Lyg} fits.  The resulting column densities are virtually
identical for non-saturated lines.  Even in the saturated higher
column density lines, the resulting column densities are highly
consistent with a one-to-one correlation.  We thus have validated that
omitting the {\Lyg} transition provides no skew in the resulting {\HI}
column densities.

\section{Deblending}
\label{sec:deblending}

In two cases, we identified absorption components blended with other
absorption features from a different redshift.  Here, we illustrate
our deblending technique.

\begin{figure*}[thb]
\epsscale{0.9}
\plotone{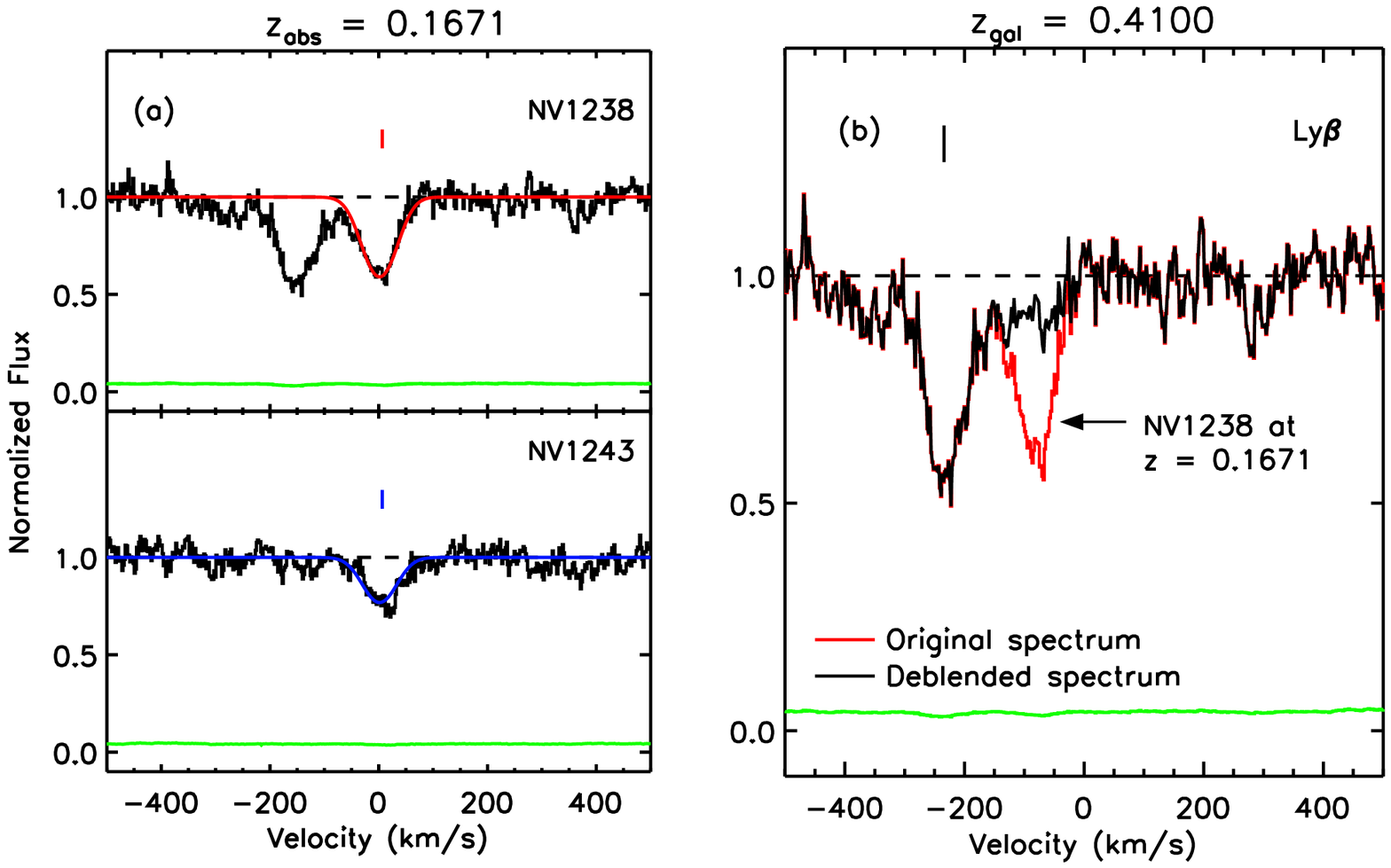}
\caption{Deblending of {\Lyb} for absorption associated with the
  galaxy at $z_{\rm gal} = 0.4100$ in the field of Q0405$-$123. The
  {\Lyb} line is blended with {\NVfirst} at $z_{\rm abs} =
  0.1671$. The red spectrum is the original data and the black
  spectrum is the result of the deblending process.}
\label{fig_Q0405deblend}
\end{figure*}

The first case occurs in the {\Lyb} line associated with the $z_{\rm
  gal} = 0.4100$ galaxy in the field of Q0405$-$123.  The {\Lyb} is
blended with {\NVfirst} at $z_{\rm abs} = 0.1670$.  In
Figure~\ref{fig_Q0405deblend}, we show the results of removing the
contaminating {\NVfirst} line. The red lines show the fit and data for
the contaminating {\NVfirst} line. In blue is the accompanying Voigt
profile fit for the corresponding {\NVsecond} line.

\begin{figure*}[hbt]
\epsscale{0.9}
\plotone{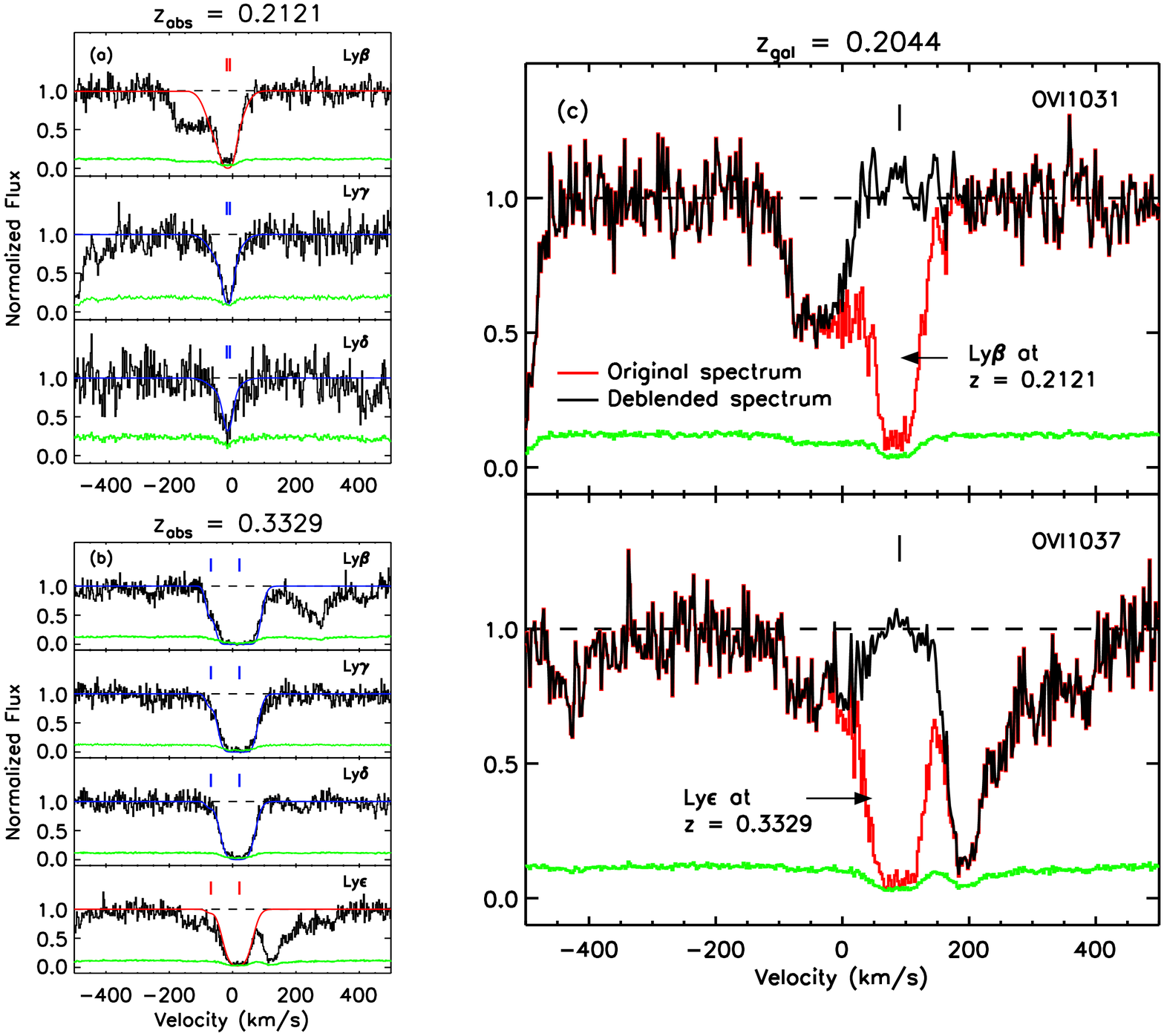}
\caption{Deblending of {\OVIdblt} for absorption associated with the
  galaxy at $z_{\rm gal} = 0.2044$ in the field of Q1136$-$1334. The
  {\OVIfirst} line is blended with {\Lyb} at $z_{\rm abs} = 0.2121$
  and the {\OVIsecond} line is blended with {\Lye} at $z_{\rm abs} =
  0.3329$.  The red spectrum is the original raw data and the black
  spectrum is the result of the deblending process.}
\label{fig_Q1136deblend}
\end{figure*}

The second case occurs in the absorption associated with the galaxy in
the field of Q1136$-$1334 at $z_{\rm gal} = 0.2044$.  Both members of
the {\OVIdblt} doublet suffer blending with Lyman-series absorption
from two different higher-redshift absorbers ({\OVI}~$\lambda 1031$ is
contaminated by {\Lyb} at $z_{\rm abs} = 0.2121$ and {\OVI}~$\lambda
1037$ is contaminated by {\Lye} at $z_{\rm abs} = 0.3329$).  In
Figure~\ref{fig_Q1136deblend}, we show the deblending results. Again,
we highlight in red both the fits and subtracted components of the
contaminating transitions ({\Lyb} and {\Lye}).

\bibliographystyle{apj}
\bibliography{bibliography}

\end{document}